\documentclass[fleqn,usenatbib]{mnras}

\usepackage[T1]{fontenc}

\DeclareRobustCommand{\VAN}[3]{#2}
\let\VANthebibliography\thebibliography
\def\thebibliography{\DeclareRobustCommand{\VAN}[3]{##3}\VANthebibliography}

\usepackage{graphicx}	% Including figure files
\usepackage{amsmath}	% Advanced maths commands
\usepackage{amssymb}	% Extra maths symbols

\usepackage{newtxtext,newtxmath}
\usepackage{graphicx}
\usepackage{siunitx}
\usepackage{mathtools}
\usepackage{mathrsfs}
\usepackage{physics}

\usepackage{ulem}
\usepackage{soul}
\usepackage[dvipsnames]{xcolor}

\newcommand{\Kepler}{\textit{Kepler}}

\title%
[%
Asteroseismology of
the dip structure in period-spacings
]
{%
Asteroseismology of
the dip structure in period-spacings
of rapidly rotating $\gamma$ Doradus stars
caused by the coupling between
core and envelope oscillations
}

\author[T. Tokuno and M.Takata]{
Takato Tokuno,$^{1}$ $^{2}$\thanks{E-mail: tokuno-takato@g.ecc.u-tokyo.ac.jp (TT)}
Masao Takata $^{1}$
\\
$^{1}$ Department of Astronomy, School of Science, The University of Tokyo, 7-3-1 Hongo, Bunkyo-ku, Tokyo 113-0033, Japan \\
$^{2}$ Graduate School of Arts \& Sciences, The University of Tokyo, 3-8-1 Komaba, Meguro, Tokyo 153-8902, Japan 
}

\date{Accepted XXX. Received YYY; in original form ZZZ}

\pubyear{2021}

\begin{document}
\label{firstpage}
\pagerange{\pageref{firstpage}--\pageref{lastpage}}
\maketitle

\begin{abstract}
Recent asteroseismic observations by the \Kepler\ space mission
have revealed
the dip fine structure in
the period-spacing versus period diagram
of  rapidly rotating $\gamma$ Doradus stars.
Following the successful 
reproduction of the dip structure by numerical calculations
in previous studies,
we present in this paper the physical mechanism
of how the dip is formed
as a result of the interaction
between the gravito-inertial waves in the radiative envelope
and the pure inertial waves in the convective core.
We analytically describe
the wave solutions 
in both of the radiative envelope
and the convective core,
and match them at the interface to construct an eigenmode.
We have found from the analysis the following points:
the dip structure is mainly controlled
by a parameter that 
has an inverse correlation with
Brunt--V\"ais\"al\"a frequency at the interface; the depth and the width of the dip
is shallower and larger, respectively,
as the parameter gets large;
the shape of the dip can be approximated by
the Lorentzian function;
the period at the central
position of the dip is equal to or slightly smaller than that of
the involved pure inertial mode in the convective core.
We 
have also understood
based on the evolutionary models of main-sequence stars
that the parameter is inversely correlated with the chemical composition gradient at the convective-core boundary.
The dip structure 
thus would provide
information about the poorly-understood physical processes,
such as diffusion, convective overshooting and rotational mixing,
around the boundary
between the convective core and the radiative envelope.
\end{abstract}


%%%%%%%%%%%%%%%%%%%%%%%%%%%%%%%%%%%%%%%%%%%%%%%%%%

%%%%%%%%%%%%%%%%% BODY OF PAPER %%%%%%%%%%%%%%%%%%

\section{Introduction}

Theory of stellar structure and evolution
is one of the cornerstones in our understanding of the universe.
It is involved in a lot of key problems in 
astrophysics,
including
nucleosynthesis (and hence chemical evolution),
galactic archaeology,
and
the dynamo mechanism to
generate and maintain the magnetic field.
The conventional approach to study stellar physics
is to compare
the observed surface properties of stars,
such as effective temperature, gravity and chemical compositions,
with
the theoretical models that are constructed
based on fundamental physical laws.
While this method has been fairly successful in describing
the global picture of stellar evolution,
there still remain some important questions.
Particularly,
we have poorly understood
the details of the physical processes 
that are related to
convection, 
rotation and the magnetic field
in the stellar interior
\citep[e.g.][]{kippenhahn2012stellar,Maeder:2009th}.

Asteroseismology is a relatively new and rapidly growing field
in stellar physics.
It aims to probe the internal structure of stars 
with the oscillations detected at the surface.
Thanks to recent space missions
such as
CoRoT \citep{Baglin:2006aa}, \Kepler\ \citep{Borucki:2010aa} and TESS \citep{Ricker:2014aa},
we have obtained
massive and high-quality seismic data
of a variety of pulsating variable stars.
These data make it possible
to study the unresolved problems in stellar physics
from a different angle.

One of the long-standing problems in 
the theory of stellar structure and evolution
is that of rotation.
It is coupled with
various types of 
hydrodynamical and magnetohydrodynamical
instabilities,
and accordingly
induces the transport and mixing of materials.
The redistribution of chemical elements
that are involved in nuclear reactions
directly affects nucleosynthesis,
the evolutionary track on the HR diagram,
and the lifetime of the stars.
It is also considered to play an essential role
in generating and maintaining the magnetic field in the stars.

We confine ourselves in this paper to the
$\gamma$ Doradus ($\gamma$ Dor) stars
\citep{1992Obs...112...53C,Balona:1994aa,1994ComAp..17..213K}.
They are main-sequence stars with
the mass of $1.4$--$2.0\,\mathrm{M}_{\odot}$
and
the typical rotation period of about $1$ day.
They
exhibit oscillations with 
multiple periods of about $1$ day \citep[e.g.][]{Li2020Gravity}, which is 
comparable to their typical rotation period.
Since the dynamical time-scale of these stars
is about a few hours,
we understand that
the oscillations are non-radial high-order gravity modes,
but with the considerable impact of rotation.
In fact, 
recent studies have demonstrated that
the main contribution to their oscillation spectrum
comes from the gravito-inertial modes,
which 
are composed of
gravity waves strongly affected by the Coriolis force
\citep{Van-Reeth:2015aa,saio2018theory,saio2018astrophysical}.
More specifically,
\citet{Li2020Gravity}
have shown that
the most commonly observed modes
in $\gamma$ Dor stars
are Kelvin g modes
(or prograde sectoral g modes),
and that
Rossby modes
\citep[or r modes;][]{Papaloizou:1978aa}
are also detected in about 20 per cent
of the stars in their sample.
Once we identify the observed eigenmodes
and understand their physical properties,
we may interpret the observed oscillation spectrum
in terms of a few structural parameters,
including
the average rotation period
heavily weighted in the deep radiative region
and the characteristic period of the oscillations
\citep{Van-Reeth:2016aa,Christophe:2018aa,Li2020Gravity,Takata:2020aa,Takata2020inferring}.

High-precision data from the \Kepler\ mission
further allow us to examine
the fine structure in
the frequency spectrum of $\gamma$ Dor stars.
\cite{saio2018astrophysical}
have noticed
in the analysis of KIC\,5608334
that
the distribution of period-spacings
deviates in a small period range
from the one expected
by
Kelvin g modes trapped in the
radiative envelope,
though they 
misinterpret that
the structure 
is caused by
the interaction with
other types of gravito-inertial modes,
which are also confined in the radiative envelope.
\cite{ouazzani2020first}
have then shown based on numerical calculations
that
the eigenmodes in the relevant period range
have finite amplitude not only in the radiative envelope,
but also in the convective core,
where the buoyancy force does not exist.
They have interpreted these modes to consist of 
the pure inertial waves in the convective core, which are restored by only the Coriolis force, and the gravito-inertial waves in the radiative envelope.
The numerical calculations
by \cite{saio2018astrophysical}
and those by \cite{ouazzani2020first}
agree that
the period-spacings are smaller 
in the relevant period range
so that their distribution follows 
a characteristic dip structure.
In fact,
\cite{saio2021rotation}
have identified
such dips in a sample of $\gamma$ Dor stars
observed by the \Kepler\ mission,
and succeeded in
constraining the rotation period of the convective core
by constructing
the evolutionary model of each star
that has consistent dip structure with the observations.

While
\cite{saio2021rotation}
have clearly demonstrated
the potential of asteroseismology
based on the dip structure
in the period-spacing distribution
to diagnose the properties of the convective core
of $\gamma$ Dor stars,
it is fair to say that
the physical mechanism to form the dip structure
is yet to be made clear in detail.
The main purpose of this paper is
to treat this problem
based on the analytical approach
rather than the numerical one
adopted by
\cite{ouazzani2020first}
and
\cite{saio2021rotation}.
By this approach,
we can elucidate 
which physical parameters are most important
for the dip formation,
and discuss how these parameters depend
on the mass and the evolutionary stage.

The outline of this paper follows.
In Section \ref{sec:two}, we first
propose the general
picture of the dip formation without any detailed analysis.
In Section \ref{sec:three}, 
we 
analytically
formulate 
the problem of eigenoscillation
by matching the pure inertial waves in the convective core
with the gravito-inertial waves in the radiative envelope.
Solving the problem approximately,
we make clear the properties of the dip structure
in the period-spacing versus period diagram.
In Section \ref{sec:four}, we compare
the derived expressions with the results
of numerical calculations
to investigate the evolutionary change of the dip structure
during the main-sequence stage
and to confirm the validity of the approximated expressions.
Discussions and conclusions are presented in Sections 
\ref{sec:five} and \ref{sec:six}, respectively.

\section{Picture of the dip formation}
\label{sec:two}

\begin{figure}
    \centering
	\includegraphics[width=\columnwidth]{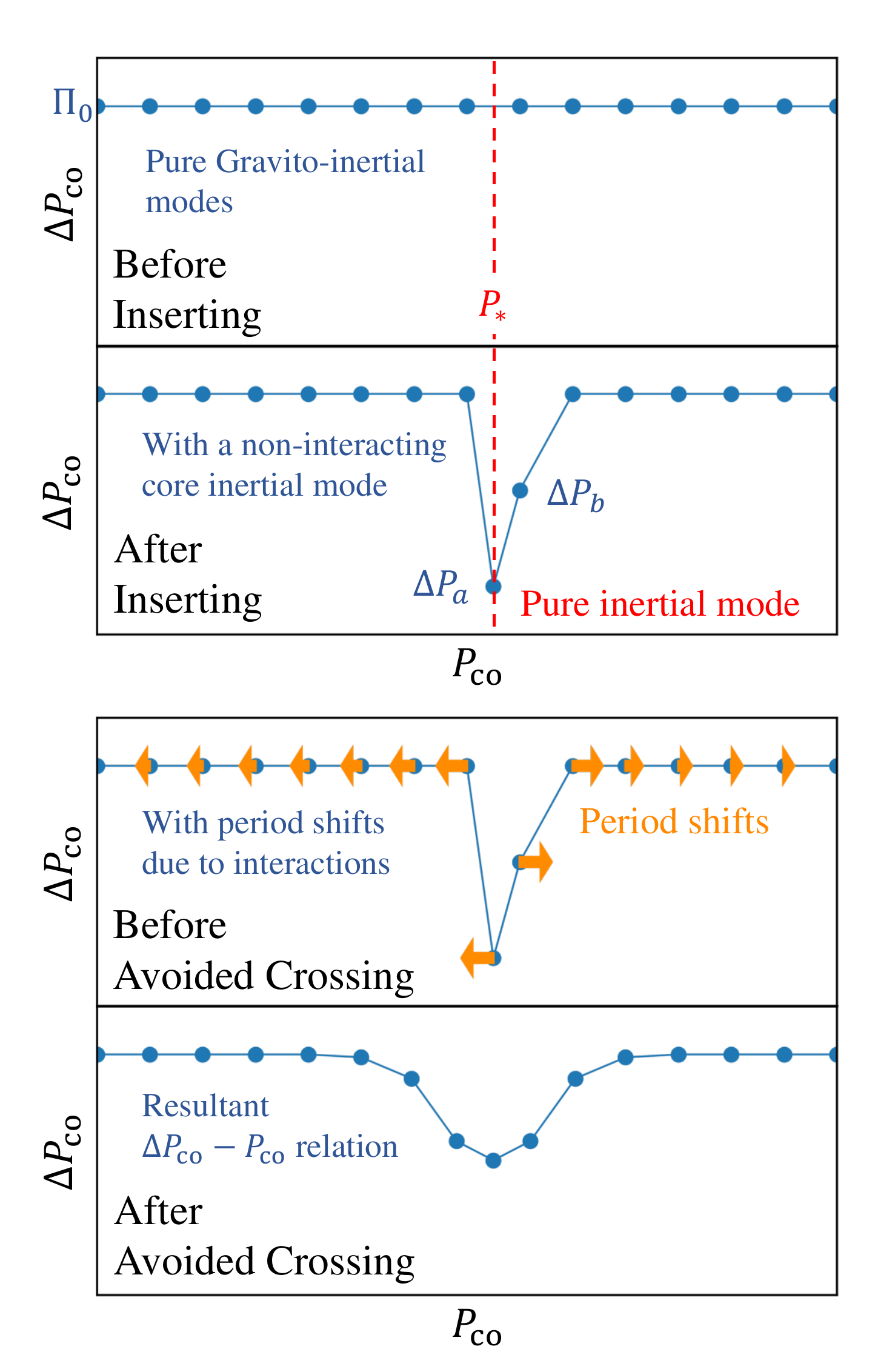}
    \caption{Schematic pictures of dip formation
    in the $P$--$\Delta P$ diagram, where the period $P$ and the period spacing
    $\Delta P$ are both measured in the co-rotating
    frame. 
    Upper panel (step 1 and 2):
    a pure-inertial mode is inserted in
    the spectrum of the constant period-spacing
    of the gravito-inertial modes.
    Lower panel (step 3):
    the dip is formed as a result of the avoided crossing.
    }
    \label{fig:picture_dip}
\end{figure}

In this section,
we propose a physical picture to explain
how the dip structure 
in a period-spacing ($\Delta P$) versus period ($P$) diagram
is formed as a result
of the interaction
between the pure-inertial waves in the convective core
and the gravito-inertial waves in the radiative envelope.
Our argument consists of three steps.

In the first step,
we consider the case where
the oscillations are completely evanescent in the convective core
irrespective of the frequency,
so that there exist only 
gravito-inertial modes trapped in the radiative envelope.
Among those modes,
we concentrate on Kelvin g
modes
(or prograde sectoral g modes)
because they have the highest visibility in a rapidly rotating star.
The oscillation periods of these modes
are (almost) equally spaced in the co-rotating frame.
Therefore, the corresponding period-spacing $\Delta P$ is constant (with a constant period-spacing of $\Pi_0$),
as shown in the top panel of Figure \ref{fig:picture_dip}.
This property of the Kelvin modes is totally analogous
to that of the high-order g modes in the absence of rotation.

We next take account of
the pure inertial modes that are confined in the convective core,
but have no interaction with the envelope oscillations.
Because the frequency spectrum of
those modes are much more sparse than the Kelvin 
g modes,
we may pay attention to only one pure-inertial mode.
If the period of the pure-inertial mode $P_*$ lies between the two consecutive periods $P_a$ and $P_b$
of the gravito-inertial modes,
the two period-spacings $\Delta P_a = P_*-P_a$  and $\Delta P_b = P_b - P_*$
are smaller than $\Pi_0$,
whereas 
the other period-spacings remain to be $\Pi_0$.
This is shown in the second panel of Figure \ref{fig:picture_dip}.

In the final step,
we consider the interaction between 
two consecutive modes.
We may regard that
the interaction causes
the mode periods to repulse each other
(cf.\ the second last panel of Figure~\ref{fig:picture_dip}).
This phenomenon is essentially the same as the avoided crossing
\citep[cf.][]{osaki1975nonradial,1977A&A....58...41A}.
Since the repulsing force should be stronger if 
the two periods
are closer to each other,
the period-spacings near $P_*$ get larger because of interaction,
whereas those far from $P_*$ are hardly changed.
The $P$--$\Delta P$ diagram is accordingly modified
so that the two small values $\Delta P_a$ and $\Delta P_b$ get slightly larger,
and other values close to them become smaller than $\Pi_0$.
We thus observe the formation of a dip
as in the bottom panel of Figure \ref{fig:picture_dip}.

Based on the presented picture,
we may derive a simple relation of the period-spacings.
For this purpose,
let us evaluate
the sum of the period-spacings 
in a wide range of the period
that includes the dip.
If the mode periods including $P_*$ in the range are
expressed by $P_n$ with
the integral indices $n$ 
in the range of $n_{\mathrm{min}} \le n \le n_{\mathrm{max}}$, 
where the smaller values of $n$ correspond to the shorter periods,
the period-spacing is defined as
\begin{equation}
\Delta_n P = P_{n+1} - P_{n},
\end{equation}
so that its sum is given by
\begin{equation}
\sum_{n=n_{\mathrm{min}}}^{n_{\mathrm{max}}-1} \Delta_n P
=
P_{n_{\mathrm{max}}}-P_{n_{\mathrm{min}}}
.
\label{eq:sumdP}
\end{equation}
Since 
the mode periods are scarcely changed by the interaction
if they are far from $P_*$,
the difference 
on the right-hand side of equation (\ref{eq:sumdP})
can be calculated
based on the period spectrum of the gravito-inertial modes,
which essentially
has 
the constant spacing of $\Pi_0$.
Because there are in total $n_\mathrm{tot}=n_{\mathrm{max}}-n_{\mathrm{min}}+1$
modes in the range,
the numbers of the gravito-inertial modes 
and their period-spacings
are $n_\mathrm{tot}-1$ and $n_\mathrm{tot}-2$, respectively.
We therefore obtain $\sum_n \Delta_n P \approx (n_\mathrm{tot}-2)\Pi_0$,
which is equivalent to

\begin{gather}
\sum_{n=n_\mathrm{min}}^{n_\mathrm{max}-1} \left( 1 - \frac{\Delta_n P}{\Pi_0} \right) \approx 1 .
\label{integrate}
\end{gather}

\section{Formulation and Analysis}
\label{sec:three}

\begin{figure}
    \centering
	\includegraphics[width=\columnwidth]{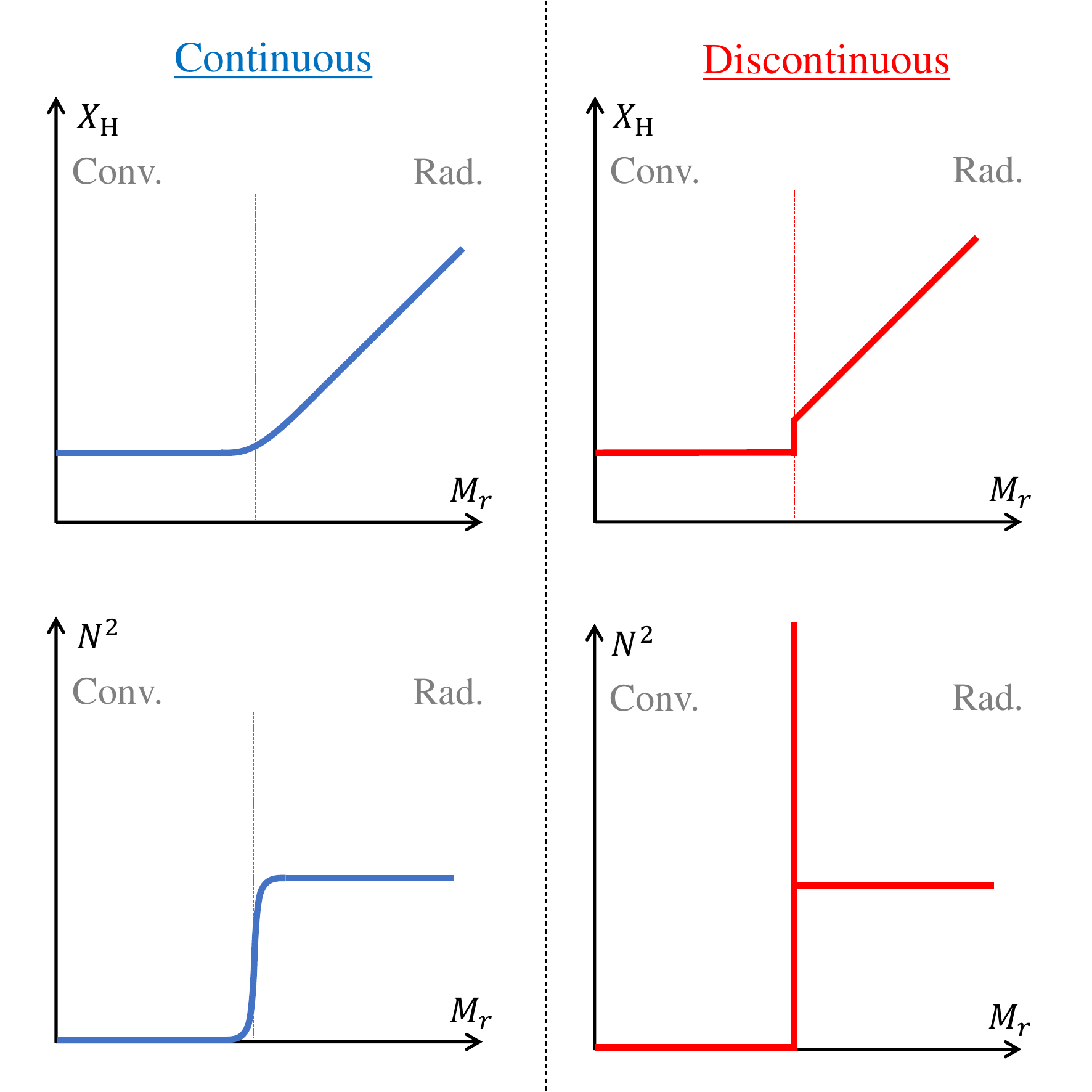}
    \caption{%
        Schematic profiles
        of the hydrogen mass fraction $X_\mathrm{H}$ and
        (squared) Brunt-V\"{a}is\"{a}l\"{a} frequency $N^2$ for the continuous (left panels) and discontinuous (right panels) cases.
        The abscissas mean the concentric mass, while the ordinates correspond to $X_\mathrm{H}$ in the upper panels
        and $N^2$ in the lower panels. The vertical dashed lines 
        represent the boundary between the
        convective core
        and the radiative envelope.%
    }
    \label{fig:Schematic_XH}
\end{figure}

This section shows our analysis about the resonance of the waves between the convective core and radiative envelope. We first formulate the expressions of oscillations in the envelope (Section \ref{subsubsec:three_one_one}) and the (uniform density) core (Section \ref{subsubsec:three_one_two}) separately. We consider the uniform density core because the analytical solutions are exceptionally known for this case.

Then, we match the oscillations in the two regions by requiring the continuity of the radial displacement $\xi_r$ and the Lagrangian pressure perturbation $\delta p$ at the boundary (Section \ref{subsec:three_two}).
Solving the matching conditions approximately, we obtain an approximate relation between the oscillation period $P$ and the period-spacing $\Delta P$ including the effect of coupling (Section \ref{subsec:three_three}). We additionally show that the analysis can formally be justified even if the core density is not uniform (Section  \ref{subsec:three_four}).

Before starting the analysis, we make a remark about the structure near the boundary between the convective core
and the radiative envelope.
While we assume up to Section \ref{subsec:three_four} that 
the profiles of the hydrogen mass fraction $X_\mathrm{H}$ and hence
the density $\rho$ are continuous at the boundary,
we then present an analysis for the case when these profiles are discontinuous
in Section \ref{subsec:three_five}. Fig ~\ref{fig:Schematic_XH} shows the schematic profiles of $X_\mathrm{H}$ and
(squared) Brunt-V\"{a}is\"{a}l\"{a} frequency $N^2$ to indicate the difference between the continuous and discontinuous cases.

From a physical point of view,
we should apply the continuous (discontinuous) case if the scale height of $N^2$ at the boundary
is much larger (smaller)
than the local wavelength
of gravito-inertial waves
around the peak of $N^2$
in the radiative envelope.
The reason why we consider the
both cases is that
the structure near the
boundary is not understood well.
We should determine which case should be applied depending the evolutionary stage
and the physical processes considered.

\subsection{Equation of oscillations }
\label{subsec:three_one}

The fundamental equations of the present analysis are those
of linear adiabatic oscillations of rotating stars.
We assume that the equilibrium structure is spherically symmetric,
which is a good approximation for low-frequency oscillations trapped mainly in the deep interior of the stars.
We also assume for simplicity
uniform rotation.

\subsubsection{Envelope oscillations}
\label{subsubsec:three_one_one}

The oscillations in the envelope of $\gamma$ Dor stars are composed of gravitio-inertial waves, of which restoring force is the buoyancy combined with the Coriolis force.
Those oscillations can be described accurately
by
the traditional approximation of rotation (TAR) \citep[e.g.][]{Eckart2015hydrodynamics,aerts2010asteroseismology}.
In this framework,
we neglect the horizontal component of the angular velocity of rotation and the perturbation to the gravitational potential (the Cowling approximation).
The most important point is that
the oscillation variables under TAR are separable in
the spherical coordinates
$(r, \theta, \phi)$
in the co-rotating frame.
The radial displacement $\xi_r$
and
the Eulerian pressure perturbation $p'$
can be written as
\begin{equation}
        \xi_{r}=\xi_{r}(r) \Theta_{k}^m(\mu ; s ) e^{i m \phi - i \omega t}  %\notag
\end{equation}
and
\begin{equation} 
        p^{\prime}=p^{\prime}(r) \Theta_{k}^m(\mu ; s ) e^{i m \phi - i \omega t},
\end{equation}
respectively.
Here, $\xi_{r}(r)$ and $p^{\prime}(r)$
stand for the parts of radial dependence.
The axial symmetry of the equilibrium structure leads to the azimuthal dependence of $e^{i m \phi}$, where $m$ is the azimuthal order.
Since we choose the time ($t$) dependence of $e^{-i\omega t}$,
positive (negative) values of $m$ correspond to prograde (retrograde) modes.
On the other hand, the $\theta$ dependence 
(through $\mu=\cos\theta$)
is described by the Hough function,
$\Theta_{k}^m(\mu; s)$, which is the eigenfunction of the Laplace tidal equation.
The integral index $k$ is introduced by 
\citet{lee1997low} to distinguish the kind of modes,
and
the spin parameter $s$ is defined by
$s = 2 \Omega / \omega$,
in which $\Omega$ is the angular velocity of rotation.

The radial parts of the variables satisfy
\begin{equation}
        \frac{{d} \xi_{r}(r)}{{d} r}=-\left(\frac{2}{r}-\frac{1}{\Gamma_{1}H_{p}} \right) \xi_{r}(r)+\frac{1}{\rho c_s^{2}}\left(\frac{\lambda c_s^{2}}{r^{2} \omega^{2}}-1\right) p^{\prime}(r), 
        \label{radius}
\end{equation}
and
\begin{equation}
        \frac{{d} p^{\prime}(r)}{{d} r}=\rho\left(\omega^{2}-N^{2}\right) \xi_{r}(r)-\frac{p^{\prime}(r)}{\Gamma_{1}H_{p}},
        \label{radius2}
\end{equation}
where $H_{p}$, $N$, $\Gamma_{1}$, $\rho$ and $c_s$ mean
the pressure scale height,
the Brunt-V\"{a}is\"{a}l\"{a} frequency,
the first adiabatic index,
the density and the sound speed,
respectively. 
Equations
(\ref{radius}) and (\ref{radius2}) have the same form as
the oscillation equations of non-rotating stars under the Cowling approximation.
Only the difference is that,
in the absence of rotation, the eigenvalue of the Laplace tidal equation, $\lambda$,
is replaced with $\ell(\ell+1)$, where $\ell$ is the spherical degree.
We can therefore analyse equations (\ref{radius}) and (\ref{radius2}) in the same way as in the case without rotation.

Because the oscillation periods of $\gamma$ Dor stars
are much longer than the dynamical time-scale of the stars,
the (radial) wavelengths of the constituent waves are short enough for the JWKB type asymptotic analysis
to be applicable
\citep[cf.][Section.~16]{unno1989nonradial}.
We first introduce new dependent variables $v$ and $w$ by
\begin{equation}
    v \equiv \rho^{1/2} c_s r^2 \left| 1- \frac{\lambda c_s^2}{r^2\omega^2}\right|^{-1/2} \frac{\xi_r(r)}{r}.
    \label{v} 
\end{equation}
and
\begin{equation}
    w \equiv \rho^{1/2} c_s^2 r \left| N^2- \omega^2 \right|^{-1/2} \frac{p'(r)}{\rho c_s^2},
    \label{w} 
\end{equation}
respectively.
We can show from equations (\ref{radius}) and (\ref{radius2}) that
$v$ and $w$ satisfy
\begin{equation}
    \frac{d^2 v}{d r^2} + k_r^2 v \simeq 0 \label{vequation}
\end{equation}
and
\begin{equation}
    \frac{d^2 w}{d r^2
    } + k_r^2 w \simeq 0, \label{wequation} 
\end{equation}
respectively,where $k_r$ denotes the wave number defined as 
\begin{gather}
    k_r^2 %=-\kappa^2 
    \equiv \frac{\omega^2 - N^2}{c_s^2} \left( 1- \frac{\lambda c_s^2}{r^2 \omega^2} \right). \label{kr}
\end{gather}

The solutions of equation (\ref{vequation}) and (\ref{wequation}) are
propagative (evanescent) if $k_r^2$ is positive (negative).
We mainly consider the case of $\lambda > 0$ in this paper,
and
assume that the radiative envelope consists of 
one big propagative region and surrounding small evanescent regions.
The inner and outer boundaries of the propagation region 
(turning points)
are denoted by
$r_a$ and $r_b$, respectively.
Practically, $r_a$ is determined by the condition of $\omega = N$, and is
almost equal to the radius of 
the convective core boundary (cf.~Fig.~\ref{fig:rf} in  Appendix \ref{App:A}).
On the other hand,
$r_b$ is fixed by $\omega = \mathrm{min}(N, L)$, where $L \equiv \sqrt{\lambda} c_s /r$ denotes the Lamb frequency, and is very close to the stellar surface.

The reason for introducing the two dependent variables rather than one
is that neither of the solutions of
equations (\ref{vequation}) and (\ref{wequation}) is
valid in the entire range.
This is because
the neglected terms on the right-hand side of equations (\ref{vequation}) and (\ref{wequation}) diverge at $r=r_b$ and $r=r_a$, respectively.
General solutions of the differential equations can be given as
\begin{align}
    v &\simeq \frac{1}{ \sqrt{|k_r|} } \left( \frac{3}{2} {\left| \int^r_{r_a} |k_r| \, dr \right|} \right)^{1/6} \notag \\
    & \hspace{30pt} \times \left[  a \, \mathrm{Ai} \left( \zeta_1 \right) + b \, \mathrm{Bi} \left( \zeta_1 \right) \right] 
    \quad \mathrm{for} \ r \ll r_b
    \label{vans}
\end{align}
and
\begin{align}
    w  &\simeq  \frac{1}{ \sqrt{|k_r|} } \left( \frac{3}{2} {\left| \int^{r_b}_{r} |k_r| \, dr \right|} \right)^{1/6} \notag \\
    & \hspace{30pt} \times \left[ c \, \mathrm{Ai} \left( \zeta_2 \right) + d \, \mathrm{Bi} \left( \zeta_2 \right) \right] 
    \quad \mathrm{for} \ r_a \ll r 
    \label{wans} .
\end{align}
Here $\mathrm{Ai} (x) $ and $\mathrm{Bi} (x)$ are the Airy functions of the first and second kind, respectively,
and $\zeta_1$ and $\zeta_2$ are 
defined as
\begin{equation}
      \zeta_1 \equiv \mathrm{sgn}(k_r^2) \left( \frac{3}{2} {\left| \int^{r}_{r_a} |k_r| \, dr \right|} \right)^{2/3}
\end{equation}
and
\begin{equation}
      \zeta_2 \equiv \mathrm{sgn}(k_r^2) \left( \frac{3}{2} {\left| \int^{r_b}_{r} |k_r| \, dr \right|} \right)^{2/3},
\end{equation}
respectively.
Arbitrary constants $a$, $b$, $c$ and $d$ 
(in equations \ref{vans} and \ref{wans})
should be constrained by the boundary conditions.
Following \citet[][]{unno1989nonradial}, we adopt the convention that
$\mathrm{Ai}(\zeta)$ and
$\mathrm{Bi}(\zeta)$ are the solutions
of
\begin{equation}
    \frac{d^2 y}{d \zeta^2} + \zeta y = 0
    .
\end{equation}

From the condition that $w$ should decay exponentially for $r \gg r_b$,
we first obtain $d=0$.
Then, matching the asymptotic form of $w$ with that of $v$ for $r_a \ll r \ll r_b$ based on $|k_r w| \simeq |dv/dr|$, we find
\begin{equation}
        a  = -c \sin B
\end{equation}
and
\begin{equation}
    b  = -c \cos B, \label{abc}
\end{equation}
where
\begin{gather}
    B \equiv \int^{r_b}_{r_a} k_r \, dr -\frac{\pi}{2} \simeq \frac{ \pi^2 s}{\Omega \Pi_0 } - \frac{\pi}{2}. \label{AB}
\end{gather}
Here $\Pi_0$ denotes the asymptotic
period-spacing of the gravito-inertial modes, which is
defined as
\begin{gather}
    \Pi_0 \simeq \frac{2\pi^2}{\sqrt{\lambda} }\left( \int_{r_a}^{r_b} N \, \frac{dr}{r} \right)^{-1}, \label{Pi_0}
\end{gather}
and the most right-hand side of equation (\ref{AB}) is derived from 
equation (\ref{kr}) 
under the approximations of $\omega \ll \sqrt{\lambda}c_s /r$ and $\omega \ll N$ 
for $r_a \ll r \ll r_b$. Now, we regard $\Pi_0$ as the constant for $s \gg 1$ because (1) integration over $r_a \leq r \leq r_b$ can essentially be interpreted as that over the whole radiative zone, and (2) the variation of $\lambda$ is small
(for Kelvin g modes).
This  assumption will be discussed  in Section \ref{subsec:five_four} and Appendix \ref{App:A}.

The final step of the analysis in this section is to derive
the expression of $\xi_r$ and $p'$ at the inner boundary of the radiative envelope, which we denote by $R_{\mathrm{core}}$.
Because $r_a$ is very close to $R_{\mathrm{core}}$, we may alternatively evaluate those variables at $r=r_a$
(This approximation will be justified in more detail at the
end of this section).
For this purpose, we use equation (\ref{kr}) to
expand $k_r^2$ near $r=r_a$ as
\begin{align}
    k_r^2 &\simeq \left[ \frac{\lambda}{r^2\omega^2}  \frac{dN^2}{dr} \right]_{r=r_a} (r-r_a)
    = \frac{\lambda s^2}{\epsilon^3 r_a^3}(r-r_a) , \label{kr2}
\end{align} 
in which we have utilised $N \simeq \omega$ and $\omega \ll \sqrt{\lambda}c_s /r$.
We have also introduced the parameter $\epsilon$ by
\begin{gather}
    \epsilon  \equiv \left( \frac{r_a}{4\Omega^2} \left.\frac{dN^2}{dr} \right|_{r=r_a} \right)^{-1/3} \label{epsilon},
\end{gather}
and assume $\epsilon \ll 1$
because $N$ steeply increases just outside
the convective core (cf.~Section \ref{sec:four}).

We can then derive from equation (\ref{kr2}) the following relations:
\begin{gather}
    \lim_{r \to r_a} \frac{1}{ \sqrt{|k_r|} } \left( \frac{3}{2} \int^{r}_{r_a} |k_r| \, dr \right)^{1/6} = \frac{\epsilon^{1/2} r_a^{1/2}}{s^{1/3} \lambda^{1/6}} , \label{limit1} \\ 
    \lim_{r \to r_a} \frac{d}{dr} \left[ \frac{1}{ \sqrt{|k_r|} } \left( \frac{3}{2} \int^{r}_{r_a} |k_r| \, dr \right)^{1/6} \right] = 0 \label{limit2}
\end{gather}
and
\begin{gather}
     \lim_{r \to r_a} \frac{d\zeta_1}{dr} = {\frac{s^{2/3} \lambda^{1/3}}{\epsilon r_a }}. \label{limit3}
\end{gather}
In addition, we rewrite equation (\ref{radius}) using equation (\ref{v}) as
\begin{gather}
    \frac{p^{\prime}(r)}{p \Gamma_1 } 
    = \frac{\omega}{c_s^2 \lambda^{1/2} \rho^{1/2}} \left[\frac{dv}{dr} - \left( \frac{1}{2}\frac{d \ln \rho}{dr} - \frac{1}{\Gamma_1} \frac{d \ln p}{dr} \right) v \right]. \label{ppgamma}
\end{gather}

From equations (\ref{v}), (\ref{vans}), (\ref{limit1})--(\ref{limit3}) and (\ref{ppgamma}),
we finally obtain
the expressions of $\xi_r$ and $p'$ at $r=r_a$
in the leading order of $\epsilon$
as
\begin{align}
        \left. \frac{ \xi_r(r) }{r} \right|_{r=r_a} &\simeq  Q  \epsilon X(s), 
        \label{xiransenv} \\
        \left. \frac{p^{\prime}(r)}{p\Gamma_1} \right|_{r=r_a} &\simeq Q  \frac{ r_a^2 \omega^2}{c_s^2} Y(s) ,
        \label{pprimeansenv}
\end{align}
respectively,
where we have introduced
\begin{align}
    X(s) &\equiv \lambda^{1/6} s^{2/3}  \sin( \frac{\pi^2 s }{\Omega \Pi_0 } -\frac{\pi}{6})
    \label{X}
\end{align}
and
\begin{align}
    Y(s) &\equiv \alpha\lambda^{-1/2} s^{4/3} \sin( \frac{\pi^2 s}{\Omega \Pi_0 } - \frac{5\pi}{6}).
    \label{Y} 
\end{align}
The dimensionless parameters $\alpha$ and $Q$ are
defined as
\begin{gather}
    \alpha \equiv \dfrac{3^{1/3}\Gamma(2/3)}{\Gamma(1/3)} \approx 0.73 , \label{dlp1} \\
    Q \equiv - \frac{c}{3^{2/3}\Gamma(2/3)} \frac{\lambda^{1/6}}{\epsilon^{1/2} \Omega r_a^{5/2} }  \left. \frac{1}{ \rho^{1/2}} \right|_{r=r_a}, \label{dlp3}
\end{gather}
respectively.
Since $c$ in equations (\ref{vans}) and (\ref{wans}) is an arbitrary constant,
$Q$ in equations (\ref{xiransenv}) and (\ref{pprimeansenv})
can be regarded as the normalisation constant.

Now we should confirm that the values of $\xi_r$ and $p'$ at $r = r_a$ are sufficiently close to those at $r=R_{\mathrm{core}}$, where they are fitted with an inertial oscillation in the core.
Using $N^2|_{r=R_\mathrm{core}}=0$, $N^2|_{r=r_a}=\omega^2$ and equation (\ref{epsilon}), we can show
\begin{gather}
    \frac{R_\mathrm{core}}{r_a} \sim 1-\frac{\epsilon^3}{s^2},
\end{gather}
and then, from the Taylor expansion, we can derive
\begin{gather}
    \frac{v|_{r=R_\mathrm{core}}}{ v|_{r=r_a} } \sim 1 + \mathcal{O} (\epsilon^2 / s^{4/3}) .
\end{gather} 
Considering $s \sim 10$, we conclude that $v$ at $r=r_a$ is different from that at $r=R_{\mathrm{core}}$ only 
in the second order of $\epsilon$.

Equations (\ref{xiransenv}) and (\ref{pprimeansenv}) are for  $\Theta_{k}^{m}\left(\mu; s\right)$ with a given $k$ ($\lambda$ at a given $s$ depends on $k$). In order to fit them with a core oscillation having a different $\theta$ dependence (but the common $e^{im\phi-i\omega t}$ dependence), we construct general solutions by linear combinations
as
\begin{gather}
    \left.  \frac{\xi_{r}}{r} \right|_{r=R_{\mathrm{core}}}
    =  \sum_k a_k \epsilon X_k(s) \Theta_{k}^m (\mu ; s ) \label{xirad}
\end{gather}
and
\begin{gather}
    \left. \frac{p'}{p \Gamma_1} \right|_{r=R_{\mathrm{core}}} =  \frac{\omega^2 R_{\mathrm{core}}^2}{c_s^2} \sum_k a_k Y_k(s) \Theta_{k}^m (\mu ; s ) , \label{pprimerad}
\end{gather}
in which the normalisation constant
$Q$ has been absorbed by
the coefficient $a_k$.

\subsubsection{Core oscillations (in the case of uniform density)}
\label{subsubsec:three_one_two}

The oscillations trapped in the convective core of $\gamma$ Dor stars are composed of inertial waves, for which the Coriolis force is the restoring force.  We formulate pure inertial modes in this section based on \citet{Wu2005origin}.

The equations of linear adiabatic oscillations
of the convective core (with $N=0$) 
in
the co-rotating frame
can be recast 
under the Cowling approximation
into a single equation of the scalar variable
$\psi$, which is defined as
\begin{equation}
    \psi \equiv
    {\frac{1}{\omega^2}
    \frac{p'}{\rho}}.
\end{equation}
If the density is uniform, the governing equation is reduced to
\begin{gather}
        \nabla^{2} \psi-s^{2} \frac{\partial^{2} \psi}{\partial z^{2}}= 0.  \label{pureinertial}
\end{gather}
This equation is separable in the ellipsoidal coordinates, and the solutions are given by the products of two associated Legendre functions 
with the same indices 
\citep{bryan1889}.
Assuming that $\xi_r$ and $p'$ depend on $t$ and $\phi$
through $e^{i m \phi - i \omega t}$,
we can show that they are related to $\psi$ by
\begin{gather}
    \xi_r=\frac{1}{1-s^{2}}\left(\pdv{r} -\frac{ms}{r} -\mu s^2 \pdv{z} \right)\psi 
    \label{xirpim} 
\end{gather}
and
\begin{gather}
    p' = p\Gamma_1 \frac{\omega^2}{c_s^2} \psi. \label{pprimepim} 
\end{gather}
From these relations, we obtain
the corresponding expressions at the outer boundary
of the core as%
\footnote{%
When we rewrite equation (\ref{xirpim})
in the ellipsoidal coordinate system,
we have corrected
equation (28) of \citet{Wu2005origin}
as
\begin{align*}
    \xi_{r} & = \frac{(1-x_{1}^{2})(1-x_{2}^{2})}{(1-s^{-2}) (x_{1}^{2}-x_{2}^{2})r} \Biggl[x_{1} \frac{\partial \psi}{\partial x_{1}} \frac{s^{-2}-x_{2}^{2}}{1-x_{2}^{2}}
\\
&\phantom{=}\mbox{}
    -x_{2} \frac{\partial \psi}{\partial x_{2}} \frac{s^{-2}-x_{1}^{2}}{1-x_{1}^{2}} + \frac{m}{s}\frac{x_{1}^{2}-x_{2}^{2}}{(1-x_{1}^{2})(1-x_{2}^{2})} \psi \Biggr] .
\end{align*}
}

\begin{gather}
    \left. \frac{ \xi_r(r) }{r} \right|_{r=R_\mathrm{core}} \propto C_{\ell}^m(1/s) P^m_{\ell}(\mu) 
    \label{xiranscore}
\end{gather}
and
\begin{gather}
    \left. \frac{p^{\prime}(r)}{p \Gamma_1} \right|_{r=R_\mathrm{core}} \propto \frac{ R_\mathrm{core}^2 \omega^2}{c_s^2} P^m_{\ell}(1/s) P^m_{\ell}(\mu), \label{pprimeanscore}
\end{gather}
where 
$P_{\ell}^{m}$ indicates the associated Legendre function, and
$C_{\ell}^{m}$ is defined as
\begin{gather}
    C_\ell^m(x) \equiv x \left( \frac{d P^m_{\ell}(x)}{d x}   +  \frac{m}{1 - x^2}  P^m_{\ell}(x)  \right) .
\end{gather}
Here we require for pure inertial modes $\xi_r = 0$ at the outer boundary of the convective core.
Then, the eigenmode condition can be expressed as
$C_{\ell}^{m}(1/s)=0$. Table 1 of \citet{ouazzani2020first} gives the eigenvalues $s=s_*$ for some values of $m$ and $\ell$.

By making linear combinations
of
equations
(\ref{xiranscore}) and (\ref{pprimeanscore})
for different values of $\ell$,
we may obtain the general solutions as

\begin{gather}
    \left.  \frac{\xi_{r}}{r} \right|_{r=R_\mathrm{core}}= \sum_\ell b_\ell C_\ell^m(1/s) \tilde P^m_{\ell}(\mu)
    \label{xicon}
\end{gather}
and
\begin{gather}
    \left. \frac{p'}{p\Gamma_1} \right|_{r=R_\mathrm{core}} = \frac{\omega^2 R_\mathrm{core}^2}{c_s^2} \sum_\ell b_\ell P^m_\ell(1/s) \tilde  P^m_\ell(\mu), \label{pprimecon}
\end{gather}
where $b_{\ell}$ are constant coefficients.
Note that
we have defined
the normalised associated Legendre function
by
\begin{equation}
\tilde P^m_\ell(x) \equiv 
\sqrt{
\frac{
(2\ell +1)!(\ell-m)!
}{ 2(\ell+m)!} } P^m_\ell(x) 
.
\end{equation}

\subsection{Matching the core and envelope solutions}
\label{subsec:three_two}

We match the solutions in the radiative envelope
with those in the convective core in this section.
Hereafter, suffixes ``$\mathrm{rad}$" and ``$\mathrm{con}$" denote variables in the radiative zone and the convective zone.

We require that the radial displacement $\xi_r$ and
the Lagrangian pressure perturbation $\delta p$ are continuous
at the boundary between the two zones.
Since we assume that the density $\rho$ and its derivative $d\rho/dr$ of the equilibrium structure are continuous at the boundary,
the continuity of $\delta p$ can be replaced with
that of the Eulerian pressure perturbation $p'$.

From equations (\ref{xirad}), (\ref{pprimerad}), (\ref{xicon}) and (\ref{pprimecon}),
we can formulate the continuity conditions
of $\xi_{r,\mathrm{con}}=\xi_{r,\mathrm{rad}}$ and  $p'_\mathrm{rad} =p'_\mathrm{con} $ at $r=R_\mathrm{core}$ 
as
\begin{gather}
    \sum_k a_k \epsilon X_k(s)  \Theta_{k}^m (\mu ; s) = \sum_\ell b_\ell C_\ell^m(1/s) \tilde P^m_{\ell}(\mu)  \label{xijunc} 
\end{gather}
and
\begin{gather}
    \sum_k a_k Y_k(s) \Theta_{k}^m (\mu ; s )  =  \sum_\ell b_\ell P^m_\ell(1/s) \tilde  P^m_\ell(\mu), \label{pprimejunc}
\end{gather}
respectively.
If we multiply the both sides of
equation (\ref{xijunc}) by
$\Theta_{k'}^m (\mu ; s )$ 
, where $k'$ is an integer, and integrate over $-1 \le \mu \le 1$,
we obtain from the orthonormality of
$\Theta_{k'}^m (\mu ; s )$
the expression of each $a_{k'}$,
which can then be substituted
into equation (\ref{pprimejunc})
to yield
\begin{gather}
    \left[ \mathcal{M}(s) - {\epsilon}\mathcal{N}(s) \right] \vec b = \vec 0 ,\label{MN}
\end{gather}
where
$\vec{b}$ is the vector
whose $\ell$-th element is equal to $b_\ell$,
and matrices 
$\mathcal{M}$
and
$\mathcal{N}$
are defined as
\begin{gather}
    [\mathcal{M}]_{k,\ell} = c_{k,\ell} Y_k(s) C_\ell^m (1/s)
\end{gather}
and
\begin{gather}
    [\mathcal{N}]_{k,\ell} = c_{k,\ell}  X_k(s) P_\ell^m(1/s)  ,
\end{gather}
respectively. Here, we have introduced 
$c_{k,\ell}$ by
\begin{gather} 
    c_{k,\ell} \equiv \int^1_{-1} \Theta_{k}^m (\mu ; s ) \tilde P^m_\ell(\mu)   \, d\mu .
    \label{def_ckl}
\end{gather}
For $\vec{b}$ to be non-trivial
($\vec{b}\ne\vec{0}$)
in equation (\ref{MN}),
we need to have
\begin{equation}
    \det \left[ \mathcal{M}(s) - {\epsilon}\mathcal{N}(s) \right] =0,
    \label{detMN}
\end{equation}
which is the condition to determine $s$
(and hence $\omega$).
Equation (\ref{MN})
can thus be regarded as
the eigenvalue problem about $s$.

It is generally not easy to solve
equation (\ref{MN}), which is
a matrix equation 
of infinite dimension. We consider first
the case of $\epsilon=0$ separately%
, which corresponds to $dN^2/dr = \infty$
at $R_{\mathrm{core}}$. Equation (\ref{detMN}) is reduced in this case to
(1) $Y_{k_0}\left(s\right)=0$
for a certain value $k_0$
or
(2) $C_{\ell_0}^{m}\left(1/s\right)=0$
for a certain value of $\ell_0$.
Case (1) corresponds to
the eigenvector $\vec{a}$ whose elements
are equal to zero except for $a_{k_0}$.
Then, we find from
equations (\ref{xijunc}) and (\ref{pprimejunc})
$b_{\ell}=0$ for all $\ell$.
The oscillation is a Kelvin g mode confined
to the radiative envelope
without being influenced from other modes.
On the other hand,
we may set in case (2)
$b_{\ell_0}=1$
and
$b_{\ell}=0$ for $\ell \ne\ell_0$.
Then, equation (\ref{pprimejunc})
can be used to derive
\begin{equation}
    a_k
    =
    \frac{c_{\ell_0, k}
    P_{\ell_0}^m\left(1/s\right)
    }{Y_k\left(s\right)}
    .
\end{equation}
Although the solution
corresponds to
a pure-inertial mode
in the convective core,
it also has finite amplitude
in the radiative envelope
(and hence at the surface of the star)
because $p' \ne 0$
at $r = R_{\mathrm{core}}$.
We note that
the solutions for $\epsilon=0$
(cases 1 and 2)
correspond to independent oscillations with
no avoided crossing as discussed
in Section \ref{sec:two}.

We thus understand that
$\epsilon$ is the parameter
to measure the
strength of the interaction
between the pure inertial modes
in the convective core
and the gravito-inertial modes
in the radiative envelope.
In order to solve equation (\ref{MN})
for $\epsilon\ne 0$,
we make some simplifying assumptions.
Firstly, 
the interaction 
between the oscillation in
the core and that in the envelope
is weak,
so that
$(0 < ) \epsilon \ll 1$.
Secondly,
we consider in the interaction
only
one core mode 
and
multiple envelope modes 
(with a single value of $k$)
whose frequencies are close to 
that of the core mode.
This is because
the frequency spectrum
of the pure inertial modes is much more sparse than that of
the gravito-inertial modes,
and
the interaction is significant
only between
the core mode
and the envelope modes
with close frequencies.

\begin{figure}
    \centering
	\includegraphics[width=\columnwidth]{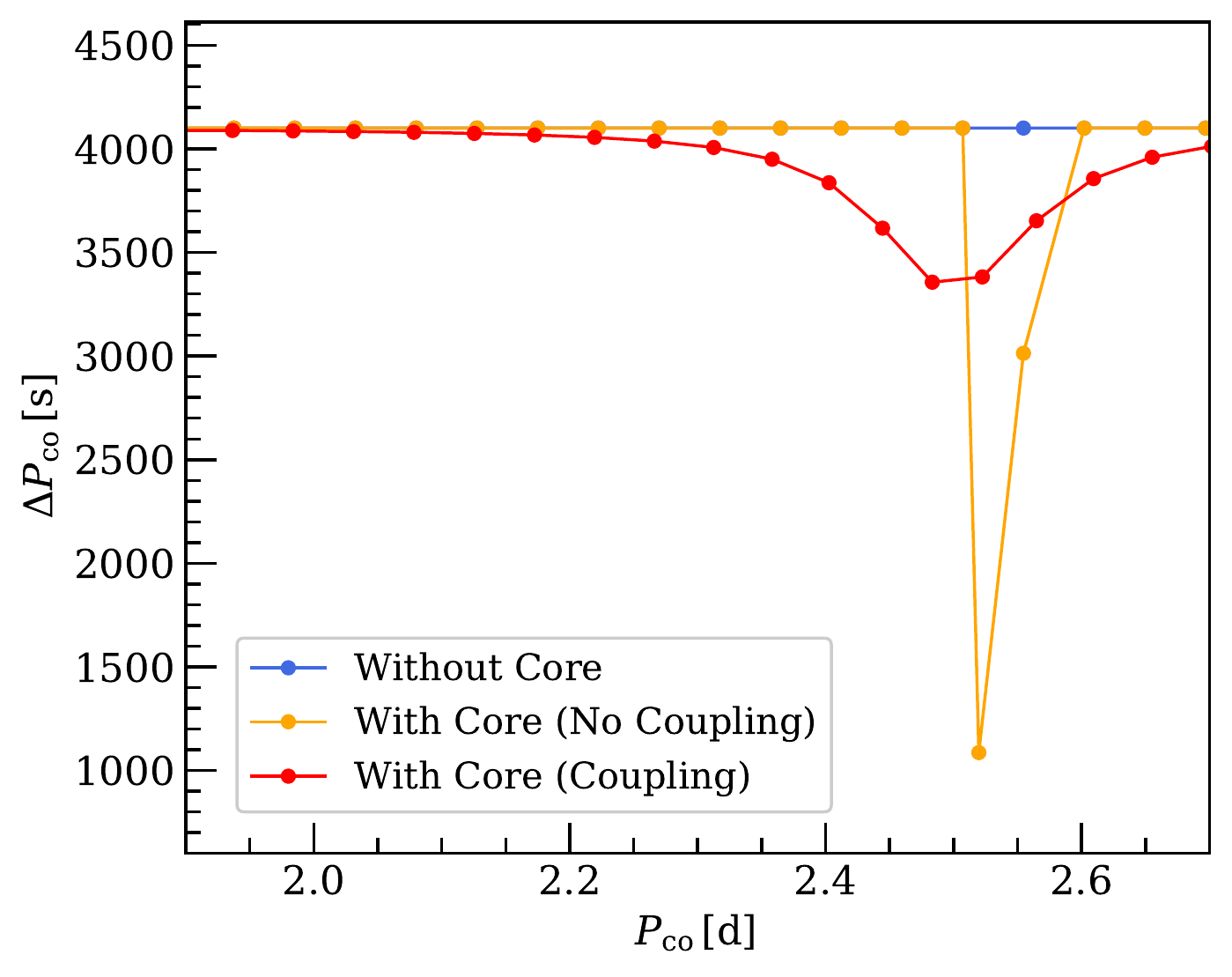}
    \caption{%
    $P$--$\Delta P$ diagram (in the co-rotating frame)
    based on the solutions of 
    equation (\ref{zeropoint_uniformdensity}). Blue, orange and red 
    points represent 
    those without the convective core,
    those with the core but
    no avoided crossing ($\epsilon=0$), and
    those with both the core and 
    avoided crossing ($\epsilon=1/100$),
    respectively. Note that we set the rotation period $2\pi / \Omega = 0.455 \, \si{d}$ and the period-spacing $\Pi_0 = 4100 \, \si{s}$.}
    \label{fig:dip_core_repulsion}
\end{figure}

Under these assumptions, we analyse equation (\ref{detMN})
in the frequency range where
the interaction is significant.
If the core mode and the envelope modes,
which are involved in the interaction,
are specified by indices $\ell$ and $k$, respectively,
we may assume that
$C_{\ell}^{m}$ and $Y_k$ are both of the order of $\epsilon$, so that
all the components in
the $\ell$-th column and the $k$-th row of
matrix $\mathcal{M} - \epsilon\mathcal{N}$
are of the order of $\epsilon$.
Since the other components 
are of the order of unity,
we can show from the cofactor expansion
that equation (\ref{detMN})
is satisfied in the leading order of $\epsilon$
if the $\left(k,\ell\right)$ component is equal to zero,
which means
\begin{gather}
    \frac{C_\ell^m (1/s)}{P_\ell^m (1/s)} {\frac{Y_k(s)}{X_k(s)}} \simeq \epsilon . \label{zeropoint}
\end{gather}
Substitution of equations (\ref{X}) and (\ref{Y}) into equation (\ref{zeropoint}) leads to
\begin{gather}
    F(s) \frac{\sqrt{3}}{2} \frac{\alpha}{\lambda^{2/3}}  s^{2/3} \left[ \cot( \frac{ \pi^2 s }{\Omega \Pi_0 }  - \frac{\pi}{6}) + \frac{1}{\sqrt{3}} \right]  \simeq \epsilon, \label{zeropoint_uniformdensity}
\end{gather}
where $F(s)$  is defined as
\begin{gather}
    F(s) \equiv -\frac{C_\ell^m (1/s)}{P_\ell^m (1/s)}.
    \label{F_uniformdensity}
\end{gather}
In Fig.~\ref{fig:dip_core_repulsion},
we plot the relation
between
the oscillation period $P$ and
the period-spacing $\Delta P$,
which are computed numerically
for $\ell=3$ \footnote{In the case of uniform density core, the pure inertial mode with $\ell=1$ is not considered because 
it corresponds to the displacement of the whole star with $\omega=0$.} and $k=0$
 (Kelvin g modes)
based on equation 
(\ref{zeropoint_uniformdensity}).
We note that $F(s)$ is given by
\begin{equation}
   F(s)
   =-\dfrac{s^2 - 10s -15}{(s+1)(s^2-5)}
   \quad
   \mbox{for $\ell=3$ and $k=0$}. \label{Fud}
\end{equation}

\subsection{Approximated solutions}
\label{subsec:three_three}

\begin{figure}
    \centering
	\includegraphics[width=\columnwidth]{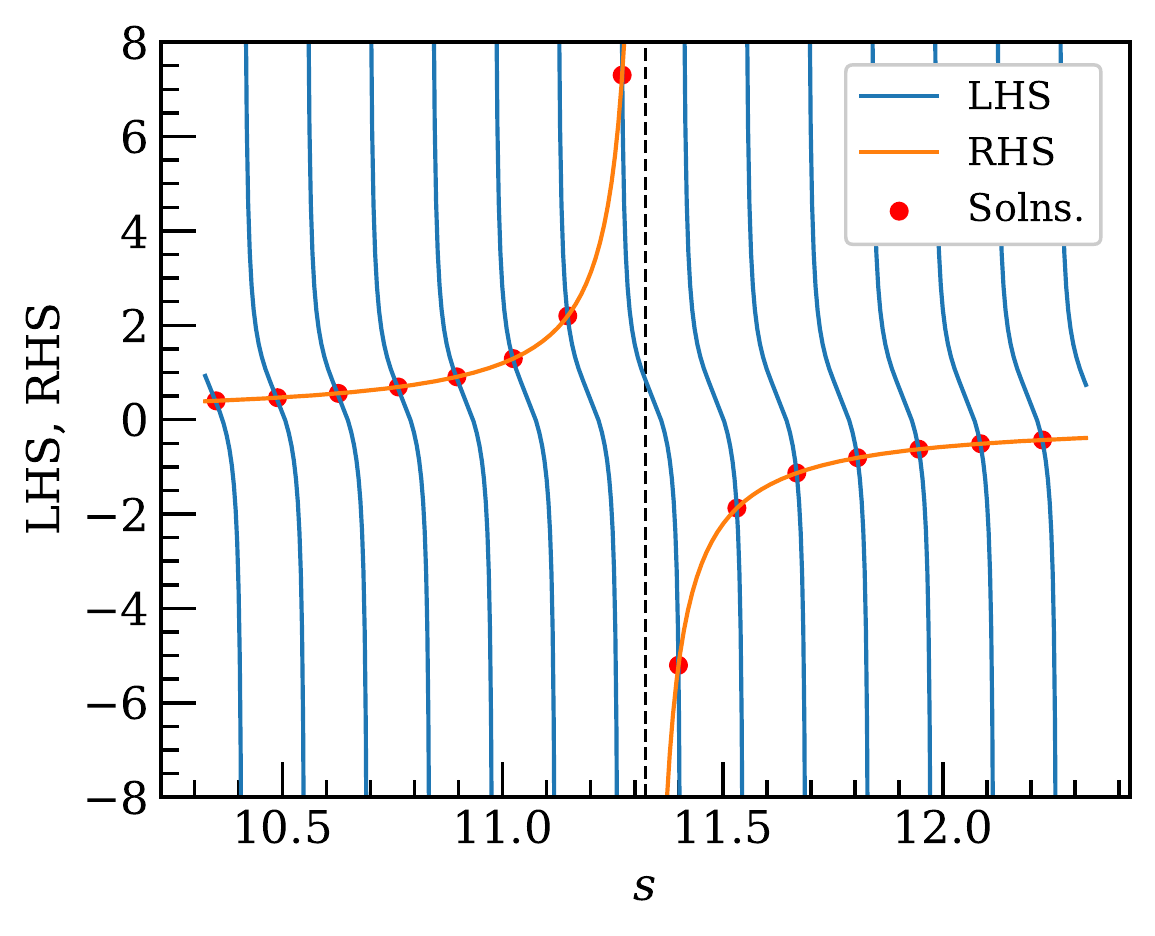}
    \caption{Solutions of equations (\ref{s12}). The horizontal axis represents spin parameter $s$. Blue and orange lines denote the left-hand side and the right-hand side of the equation, respectively. Red points represent the points of intersection between them. Black vertical dashed line indicates $s=s_*$.}
    \label{fig:zeropoint}
\end{figure}

In order to derive
an approximate analytic expression
of the period-spacing $\Delta P$
in the co-rotating frame,
we may further rewrite equation
(\ref{zeropoint_uniformdensity})
by expanding $F(s)$ near its zero point $s=s_*$, which corresponds to the spin parameter of a pure inertial mode in the core, up to the linear term in $s$.
We thus obtain
\begin{gather}
        - V(s - s_*)\left[ \cot(\frac{ \pi^2 s }{\Omega \Pi_0 } - \frac{\pi}{6}) +\frac{1}{\sqrt{3}} \right] \simeq \epsilon  \label{pict} ,
\end{gather}
where
\begin{gather}
    V \equiv \left[ - \frac{dF(s)}{ds} \frac{\sqrt{3}}{2} \frac{\alpha s^{2/3}}{\lambda^{2/3}} \right]_{s=s_*} .
    \label{V_def}
\end{gather}
Considering $s_1$ and $s_2$ as two neighbouring solutions of equation (\ref{pict}) with $s_1 > s_2$, we obtain 
\begin{align}
    \cot(\frac{ \pi^2 s_i }{\Omega \Pi_0 }  - \frac{\pi}{6}) +\frac{1}{\sqrt{3}} & \simeq - \frac{\epsilon / V}{s_i - s_*} 
\qquad    (i=1, 2)
    , \label{s12} 
\end{align}
Because the intersections 
between the left-hand side and the right-hand side
with $s=s_1$ and $s_2$ are located on
two adjacent branches of the cotangent function 
\footnote{%
In the exceptional case of $s_1 > s_* > s_2$,
$s_1$ and $s_2$ belong to the same branch of the cotangent function.
We can however obtain the same equation as equation (\ref{dcot})
even in this case
by considering the reciprocal of the cotangent function
(namely, the tangent function),
for which $s_1$ and $s_2$ are on the 
two different adjacent branches.}
(see Fig.~\ref{fig:zeropoint})%
, we premise $  \pi^2 (s_1 - s_2) / \Omega \Pi_0 - \pi \ll 1$%
, which physically means that the frequency spectrum of the envelope modes is very dense. Then, by Taylor expansion, we acquire
\begin{gather}
    \left[ 1 + \left( \frac{\epsilon / V}{
    \bar{s} - s_*} + \frac{1}{\sqrt{3}} \right) ^2 \right] \left[ \frac{ \pi^2  (s_1-s_2)}{\Omega \Pi_0 } - \pi \right] \simeq - \frac{\epsilon}{V} \frac{s_1-s_2}{(\bar{s} - s_*)^2},
    \label{dcot}
\end{gather}
in which $\bar{s}=(s_1 + s_2)/2$.
Rewriting equation (\ref{dcot})
with $P=\pi \bar{s}/\Omega$
, $\Delta P =  \pi (s_1-s_2) / \Omega$, $P_* = \pi s_* / \Omega$ and a new parameter
\begin{align}
    \sigma &\equiv \frac{3\pi \epsilon}{4\Omega V} 
    %\notag \\& =  \left( \left. \frac{dF(s)}{ds} \right|_{s=s_*} \right)^{-1}  \frac{\sqrt{3}\pi \lambda^{2/3}}{2\alpha \Omega} \left( \frac{R_\mathrm{core}P_*^2}{4\pi^2} \left.\frac{dN^2}{dr} \right|_{r=r_a} \right)^{-1/3} 
    \label{sigma}
    ,
\end{align}
we are led to the following simple 
relation
between $P$ and $\Delta P$:
\begin{gather}
    \frac{1}{\Delta P} - \frac{1}{\Pi_0} \simeq 
    \frac{\sigma / \pi }{(P - P_* + \sigma / \sqrt{3} )^2 + \sigma^2}. \label{Lorentz}
\end{gather}
The right hand of 
equation (\ref{Lorentz}) is a Lorentzian function. This indicates
\begin{gather}
    \int_{-\infty}^{\infty} \left( \frac{1}{\Delta P} - \frac{1}{\Pi_0} \right) \, dP = 1,
\end{gather}
which corresponds to
the continuity limit of equation (\ref{integrate}).

If the control parameter $\sigma$,
which is proportional to $\epsilon$,
is larger,
the width and the height
of the Lorentzian function are
larger and lower, respectively.
Equation (\ref{Lorentz}) implies that
the dip structure in the
$P$--$\Delta P$ relation
can also
be described by the (unnormalised)
Lorentzian function 
as 
\begin{gather}
    \Delta P \simeq \Pi_0 \left( 1 - \frac{\Pi_0 \sigma / \pi }{(P - P_* + \sigma / \sqrt{3} )^2 + \sigma^2 + \Pi_0 \sigma / \pi} \right) .
    \label{Lorentz2}
\end{gather}
We reemphasize that equations (\ref{Lorentz}) and (\ref{Lorentz2}) are expressions in 
the co-rotating frame of reference. We can easily show that
the width and the depth of the dip
are larger and shallower, respectively,
for larger $\sigma$ ($\epsilon$).

It should also be noted that
the Lorentzian function on the right-hand side of
equation (\ref{Lorentz}) (or equation \ref{Lorentz2})
takes its
maximum (minimum)
not
at $P=P_*$,
but at $P=P_* - \sigma/\sqrt{3}$.
Accordingly,
the period-spacing $\Delta P$ is
the smallest at a period
slightly shorter than $P_*$,
which corresponds to
the period of the pure inertial mode.

\begin{figure}
    \centering
	\includegraphics[width=1.0\columnwidth]{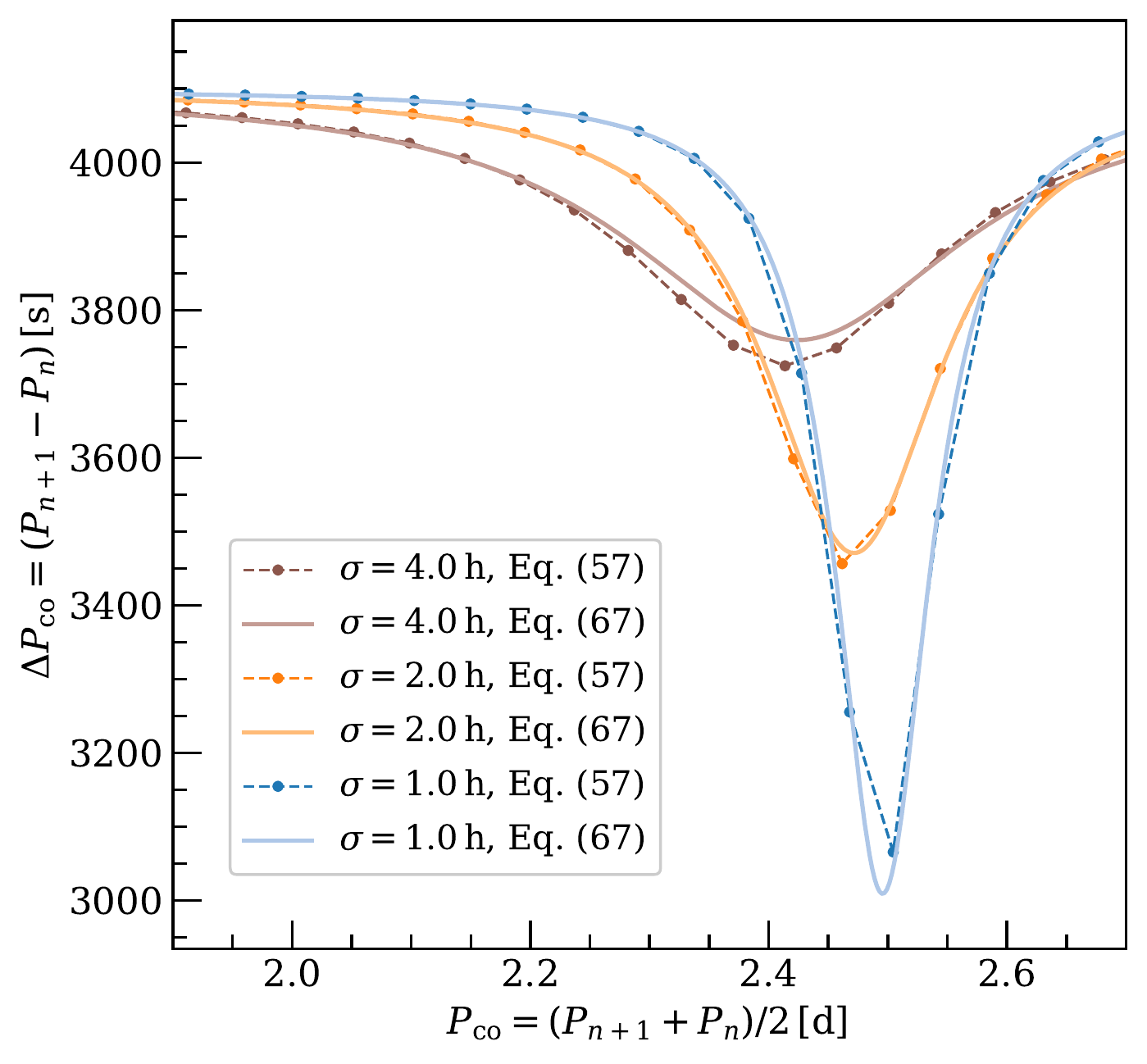}
    \caption{%
    $P$--$\Delta P$ diagram (in the co-rotating frame) with the solutions of equation (\ref{zeropoint_uniformdensity}) (filled circles connected with dashed lines) 
    and the profiles given by equation (\ref{Lorentz2}) (solid lines)
    for $2\pi/ \Omega = 0.445 \, \si{d}$, $\Pi_0 = 4100 \, \si{s}$ and
    $\sigma = 1.0$, $2.0$ and $4.0 \, \mathrm{h}$,
    which correspond to 
    $\epsilon = 6.43\times 10^{-3}$, $1.29\times 10^{-2}$
    and $2.57\times 10^{-2}$,
    respectively. We adopt equation (\ref{Fud}) for $F(s)$. Note that $P_n$ indicates the $n$-th shortest period 
    and  that the abscissa means the average period of
    two neighbouring solutions, $P = (P_n + P_{n+1}) / 2 $.
    }
    \label{fig:dip_lorentz}
\end{figure}

In Fig.~\ref{fig:dip_lorentz},
we compare the solutions of
equation (\ref{zeropoint_uniformdensity})
with the profiles computed by equation (\ref{Lorentz2}).
We generally find good agreement
up to the order of $\epsilon$.
In fact,
the small differences 
for 
$\sigma = 4.0$ hrs
are of the order of $\epsilon^2$.

\subsection{The case of non-Uniform density core}
\label{subsec:three_four}

We show in this section
that the analyses developed in
Sections
\ref{subsec:three_two}
and
\ref{subsec:three_three}
can be generalised to
the more realistic case in which
the density of the convective core
is not uniform.

For this purpose,
we formally replace
equations (\ref{xicon})
and (\ref{pprimecon})
with

\begin{gather}
    \left.  \frac{\xi_{r,\mathrm{con}}}{R_\mathrm{core}} \right|_{r=R_\mathrm{core}}= \sum_\ell g_\ell(s) \tilde P^m_{\ell}(\mu)  \label{xinud}
\end{gather}
and
\begin{gather}
    \left. \frac{p'_\mathrm{con}}{p\Gamma_1} \right|_{r=R_\mathrm{core}} = \frac{\omega^2 R_\mathrm{core}^2}{c_s^2} \sum_\ell h_\ell(s) \tilde P^m_\ell(\mu),  \label{pprimenud}
\end{gather}
respectively,
in which $g_{\ell}$ and $h_{\ell}$
can be regarded as the coefficients
of the Legendre expansion
of $\xi_{r,\mathrm{core}}$ and $p'_{\mathrm{core}}$
at $r=R_{\mathrm{core}}$, respectively.

Accordingly, we obtain instead of
equations (\ref{xijunc}) and (\ref{pprimejunc})
\begin{gather}
    \sum_k a_k \epsilon X_k(s)  \Theta_{k}^m (\mu ; s) = \sum_j b_j \sum_\ell 
    {g_{j,\ell}}(s) \tilde P^m_{\ell}(\mu)  \label{xinudjunc}
\end{gather}
and
\begin{gather}
    \sum_k a_k Y_k(s) \Theta_{k}^m (\mu ; s )  =  \sum_j b_j \sum_\ell {h_{j,\ell}}(s) \tilde  P^m_\ell(\mu), \label{pprimenudjunc}
\end{gather}
respectively, where the suffix $j$ represents the index of pure inertial modes.
Similarly to the analysis in Section \ref{subsec:three_two}, we can obtain the expression that corresponds to equation (\ref{zeropoint_uniformdensity}) as
\begin{gather}
    \tilde F(s) \frac{\sqrt{3}}{2} \frac{\alpha  s^{2/3}}{\lambda^{2/3}}  \left[ \cot( \frac{ \pi^2 s }{\Omega \Pi_0 } - \frac{\pi}{6}) + \frac{1}{\sqrt{3}} \right]\simeq \epsilon, \label{zeropoint_nonuniformdensity}
\end{gather}
where $F(s)$ in equation
(\ref{zeropoint_uniformdensity})
has been replaced by
\begin{gather}
    \tilde F(s) \equiv - \left(\sum_{\ell'} c_{k,\ell'} g_{j,\ell'}(s) \right) \left( \sum_{\ell''} c_{k,\ell''} h_{j,\ell''}(s) \right)^{-1}. \label{tildeF}
\end{gather}
We are therefore led to
equation (\ref{Lorentz}) again
with $F$ replaced by $\tilde{F}$ in
equation (\ref{V_def}).
Note that
we cannot generally write down
the analytical expressions
of $\tilde{F}$ and its
zero point $s_*$,
both of which depend
on the structure of the convective core.

\subsection{The case of non-Uniform and 
discontinuous
density profile}
\label{subsec:three_five}
We have so far assumed that both $\rho$ and $d \rho / dr$ are continuous near the core boundary. However, the numerical calculations 
suggest it more appropriate to regard that the density is discontinuous or indifferentiable  near the boundary (see Appendix \ref{App:eps_tilde}). 
We therefore present the analysis for such discontinuous cases in this section.
We suppose that the discontinuous point of the density profile is located at $r=R_\mathrm{core}$ and that the left-hand and right-hand
limits of the density are defined as
\begin{gather}
    \lim_{r \to R_\mathrm{core}-0 } \rho =  \rho_\mathrm{b} \label{rhob-}
\end{gather}
and
\begin{gather}
    \lim_{r \to R_\mathrm{core}+0 } \rho = \rho_\mathrm{b} + \Delta \rho, \label{rhob+}
\end{gather}
respectively. Note that $\Delta\rho$ is usually negative.

Accordingly, the discontinuity in
$N$ near the boundary is
generally described as
\begin{gather}
     \lim_{r \to R_\mathrm{core}-0 } N = 0 , \label{Nb-} \\
    \left. N \right|_{r=R_\mathrm{core}} = \infty \label{Ninf}
\end{gather}
and
\begin{gather}
     \lim_{r \to R_\mathrm{core}+0 } N = N_0 . \label{Nb+}
\end{gather}
The numerical results actually demonstrate
two different kinds of discontinuous cases.
In case (1), the density is continuous while its derivative is not. On the other hand, both the density and its derivative are discontinuous in case (2). In above equations (\ref{rhob-})--(\ref{rhob+}) and (\ref{Nb-})--(\ref{Nb+}), the cases (1) and (2) can be specified by setting
$\Delta \rho=0$ and $\Delta \rho \ne 0$, respectively. In addition, equation (\ref{Ninf}) is not relevant for case (1).

As in 
Section \ref{subsec:three_two}, the eigenvalue conditions can be obtained by 
matching $\xi_r$ and $\delta p$ at the boundary
between the core and envelope. While we may basically follow the same
method as in Section \ref{subsubsec:three_one_one}
to construct the envelope solutions, we can replace the Airy functions with their asymptotic forms because $k_r$ remains to be large even near the inner boundary of the propagative cavity, which coincides with the discontinuous point. With $d=0$, equations (\ref{vans}) and (\ref{wans}) 
can be rewritten into
\begin{gather}
    v \simeq -\frac{c}{\sqrt{\pi}}\frac{1}{\sqrt{k_r}} \sin \left(\int^{r_b}_{r} k_r dr -\frac{\pi}{4} \right) \label{vasym}
\end{gather}
and
\begin{gather}
    w \simeq \frac{c}{\sqrt{\pi}}\frac{1}{\sqrt{k_r}} \cos \left(\int^{r_b}_{r} k_r dr -\frac{\pi}{4} \right), \label{wasym}
\end{gather}
respectively,
and then, from equation (\ref{v}) and (\ref{w}), we obtain
\begin{gather}
    \lim_{r \to R_\mathrm{core}+0 } \frac{\xi_{r,\mathrm{rad}}}{r} 
    =  \sum_k a_k \tilde{\epsilon} \tilde{X}_k(s) \Theta_{k}^m (\mu ; s ) \label{xirans_noncont}
\end{gather}
and
\begin{gather}
    \lim_{r \to R_\mathrm{core}+0 }  \frac{p_\mathrm{rad}'}{p \Gamma_1} =  \frac{\omega^2 R_{\mathrm{core}}^2}{c_s^2} \sum_k a_k \tilde{Y}_k(s) \Theta_{k}^m (\mu ; s ), \label{pprimeans_noncont}
\end{gather}
where
\begin{gather}
    \tilde{X}_k(s) \equiv 2\lambda^{1/4} s^{1/2} \sin \left( \frac{\pi^2 s}{\Omega \Pi_0}-\frac{\pi}{4} \right)
\end{gather}
and
\begin{gather}
    \tilde{Y}_k(s) \equiv - \lambda^{-1/4} s^{3/2} \cos \left( \frac{\pi^2 s}{\Omega \Pi_0}-\frac{\pi}{4} \right).
\end{gather}
Note that the parameter $\tilde{\epsilon}$ is defined as 
\begin{gather}
    \tilde{\epsilon} \equiv \frac{\Omega}{N_0}. \label{tildeepsilon}
\end{gather}
On the other hand, the core solutions are the 
same as equations (\ref{xinud}) and (\ref{pprimenud}).

Using equations (\ref{xinud}) and (\ref{xirans_noncont}), we obtain the matching condition of $\xi_r$ as 
\begin{gather}
    \sum_k a_k 
    \tilde{\epsilon}
    \tilde{X}_k(s)  \Theta_{k}^m (\mu ; s) = \sum_j b_j \sum_\ell 
    {g_{j,\ell}}(s) \tilde P^m_{\ell}(\mu). \label{matcondxir_noncont}
\end{gather}
Similarly, from equations (\ref{pprimenud}), (\ref{pprimeans_noncont}) and (\ref{matcondxir_noncont}), 
we can derive
the condition of $\delta p =  p' + \xi_r ({\partial p}/{\partial r})$ as
\begin{align}
& \sum_k a_k \left(1+\frac{\Delta \rho}{\rho_\mathrm{b}} \right) \left[  \tilde{Y}_k(s) - \frac{GM_\mathrm{core}}{  \omega^2 R_\mathrm{core}^3 }  \tilde{\epsilon} \tilde{X}_k(s) \right] \Theta_{k}^m (\mu ; s )  \notag \\ 
    & \hspace{10pt} =  \sum_j b_j \left[  \sum_\ell h_\ell(s) - \frac{GM_\mathrm{core}}{  \omega^2 R_\mathrm{core}^3 } \sum_\ell 
    {g_{j,\ell}}(s) \right] \tilde P^m_{\ell}(\mu). 
\end{align}
With the same assumptions as 
those in Section \ref{subsec:three_two}, 
we can reduce these matching conditions to
\begin{gather}
     \cot(\frac{ \pi^2 s }{\Omega \Pi_0 } - \frac{\pi}{4}) + \tilde{\epsilon} \frac{GM_\mathrm{core}}{ \omega^2 R_\mathrm{core}^3 } \frac{\Delta \rho}{\rho_\mathrm{b}+\Delta \rho} \left. \frac{2\lambda^{1/2}}{s} \right|_{s=s*} \simeq -\frac{\tilde{\epsilon} / \tilde{V} }{(s - s_*) }  ,
     \label{eq:match_cond_disc}
\end{gather}
where
\begin{gather}
    \tilde{V} \equiv - \left( 1+ \frac{\Delta \rho}{\rho_\mathrm{b} } \right) \left[ \frac{d \tilde{F}(s)}{ds} \frac{ s}{2\lambda^{1/2}} \right]_{s=s_*} . \label{Vtilde_def}
\end{gather}
Note that $\tilde{F}(s)$ is the same  as equation (\ref{tildeF}). Equation 
(\ref{eq:match_cond_disc}) can approximately be solved for the period-spacing $\Delta P$ to yield
\begin{gather}
    \frac{1}{\Delta P} - \frac{1}{\Pi_0} \simeq 
    \frac{ \tilde{\sigma} / \pi }{(P - P_*)^2 + \tilde{\sigma}^2}, \label{Lorentz_noncont}
\end{gather}
where
\begin{gather}
    \tilde{\sigma} \equiv \frac{\pi \tilde{\epsilon}}{\Omega \tilde{V}}. \label{sigmatilde}
\end{gather}
Comparing
equation (\ref{Lorentz_noncont}) 
for the discontinuous $N^2$ profiles
with equation
(\ref{Lorentz})
for the continuous profiles,
we find the following points:
(1) 
the inverse of the period-spacing
$1/\Delta P$ 
is essentially described
by the normalised Lorentzian functions
of the period $P$
in the both cases;
(2)
the width ($\sigma$) of the Lorentzian profiles
depends on the structure of
$N^2$ near the interface 
through
$\epsilon$ 
(cf.\ equation \ref{epsilon})
and 
the property of the core inertial oscillation
through
$V$
(cf.\ equation \ref{V_def})
for the continuous case,
while
they are replaced by
$\tilde{\epsilon}$
(cf.\ equation \ref{tildeepsilon})
and
$\tilde{V}$
(cf.\ equation \ref{Vtilde_def}),
respectively,
for the discontinuous case;
(3)
the central position of the
Lorentzian profiles is located at
the period of the core inertial mode
($P_*$) for the discontinuous case,
whereas it is offset by
$-\sigma/\sqrt{3}$
for the continuous case.
It should also be noted that
the formal limit of
$dN^2/dr \rightarrow \infty$
($\epsilon\rightarrow 0$)
in the continuous case
corresponds to
the discontinuous case
with $N_0\rightarrow\infty$
($\tilde{\epsilon}\rightarrow 0$).
In this limit, 
there is no interaction of oscillations
between the convective core
and the radiative envelope.

\section{Comparison with Numerical Solutions}
\label{sec:four}

In this section,
we compare
the analyses developed
in Section \ref{sec:three} 
with the results of numerical calculations
in two respects.
We first check the main conclusion
that
the dip structure is described by
the Lorentzian profile
in Section \ref{subsec:four_two}
based on the results by
\citet{saio2021rotation}.
We then study the evolution of
the main parameter $\tilde{\epsilon}$
that controls the dip structure
in Section \ref{subsec:four_one}
based on the \textsc{mesa} stellar evolution code \citep{paxton2011modules,paxton2013modules,paxton2015modules,paxton2018modules,paxton2019modules}.
 
\subsection{Fitting the Lorentzian functions}
\label{subsec:four_two}

\begin{figure*}
	\includegraphics[width=2\columnwidth]{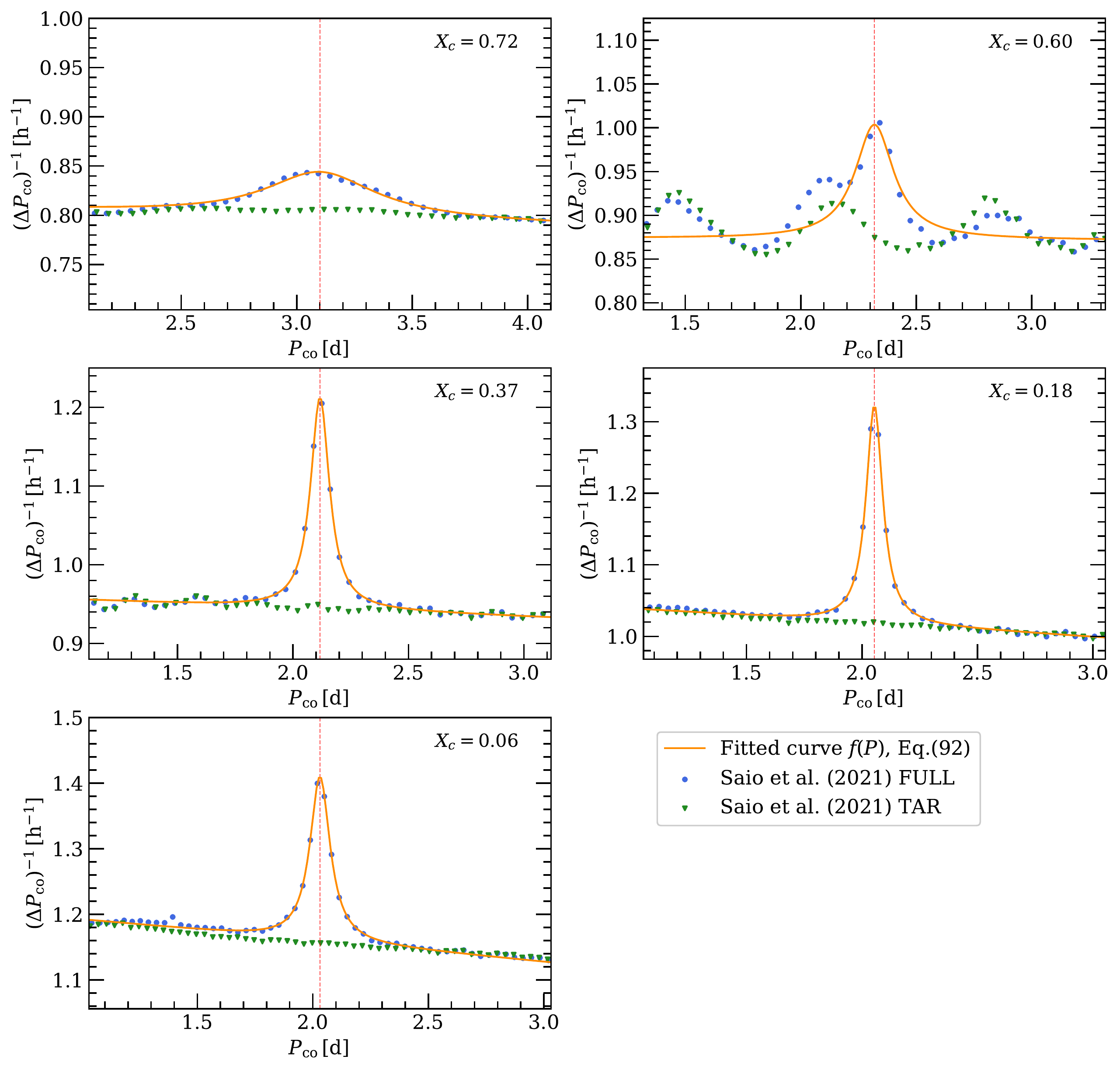}
    \caption{%
    $P$--$1/\Delta P$ diagram (in the co-rotating frame) constructed from
    the numerical results used in Fig.~4 of \citet{saio2021rotation} 
    with an additional case 
    with the central hydrogen mass fraction of $0.06$ (blue filled circles mean those of full calculation, while green triangles indicate those under the traditional approximation
    of rotation) and fitted curves by function $f(P)$, which is given by equation (\ref{fitfunction}) (orange lines).
    The stellar mass and rotation period are set to $1.5 \, \mathrm{M_\odot}$ and $0.455 \, \mathrm{d}$, respectively. In each panel, $X_c$ is the hydrogen mass fraction at the centre, which monotonically decreases with the evolution.
    The red vertical dashed lines represent the peak position of the Lorentzian function, $P_*$.}
    \label{fig:fitting_numerical}
\end{figure*}

\begin{table*}
	\centering
	\caption{Fitting parameters $x_1$, $x_2$,
	$\tilde{\sigma}$ and $P_*$
	obtained by the least-squares method in Fig.~\ref{fig:fitting_numerical} and  $\tilde{\epsilon}$ calculated 
	based on the evolutionary stellar models 
	used in Fig.~4 of \citet{saio2021rotation}.
	}
	\label{tab:fitting_result}
	\renewcommand{\arraystretch}{1.4}
	\begin{tabular}{ccccccc} 
		\hline
		$X_c$ & $x_1 \, [\mathrm{h^{-1}}]$&$x_2 \, [10^{-3} \, \mathrm{h^{-2}}]$ & 
		$(\Pi_0^*)^{-1}$
		\,$[\mathrm{h^{-1}}]$
		& $\tilde{\sigma} \, [\mathrm{h}]$ & $P_* \, [\mathrm{d}]$ & $\tilde{\epsilon}$ \\ 
		\hline 
		0.72 & $0.820 \pm 0.001 $ & $0.292 \pm 0.011$ & 
		$0.798 \pm 0.002 $ & $6.906 \pm 0.194$ & $3.101 \pm 0.008$ & 0.343 
		\\
		0.60 & $0.875 \pm 0.011$ & $0.048 \pm 0.316$ & 
		$0.873 \pm 0.029 $ &$2.428 \pm 0.184$ & $2.319 \pm 0.007$ & 0.037 
		\\
		0.37 & $0.968 \pm 0.002$ & $0.465 \pm 0.032$ & 
		$0.944 \pm 0.003 $ &$1.183 \pm 0.013$ & $2.117 \pm 0.001$ & 0.018  
		\\ 
		0.18 & $1.058 \pm 0.001$ & $0.825 \pm 0.024$& 
		$1.018 \pm 0.003$ &$1.040 \pm0.009$ & $2.054 \pm 0.000$ & 0.018 
		\\
		0.06 & $1.224 \pm 0.002$ & $1.348 \pm 0.041$& 
		$1.159 \pm 0.004$ &$1.264 \pm0.009$ & $2.031 \pm 0.000$ & 0.020
		\\
		\hline
	\end{tabular}
\end{table*}

According to equation (\ref{Lorentz}) or (\ref{Lorentz_noncont}), the inverse of the period-spacing $1/\Delta P$ is described by the Lorentzian function. We check how accurate this description is based on the numerical results given in Fig.~4 of \citet{saio2021rotation}
, which are supplemented by one more case with
the central hydrogen mass fraction ($X_c$) of $0.06$.
By carefully examining $N^2$
of the equilibrium models,
we adopt equation (\ref{Lorentz_noncont})
for the discontinuous profiles
to interpret the fitted parameters of
the Lorentzian functions.

When we analyse the numerical results,
we should note that there are two points that are not considered in deriving equation (\ref{Lorentz_noncont}).
As we see in Fig.~4 of \citet{saio2021rotation},
the period-spacing $\Delta P$ 
can have a significant
wavy component, which is generally caused by
rapid change in the equilibrium structure
\citep[e.g.][]{miglio2008probing}.
In addition, while we assume that $\Pi_0$ is constant
as a function of the period $P$,
this is not completely
supported by the numerical results of
\citet{saio2021rotation},
because the profiles of $\Delta P$ outside the dips are
inclined.
Taking these points into account,
we fit $1/\Delta P$ with function $f(P)$ defined as
\begin{gather}
    f(P) = x_1 - x_2P + 
    \frac{\tilde{\sigma} / \pi }{(P-P_*)^2 + \tilde{\sigma}^2}
    , \label{fitfunction}
\end{gather}
where 
$x_1$ and $x_2$ (as well as $\tilde{\sigma}$
and $P_*$)
are the constant parameters
to be fitted. 
Comparing $f(P)$ with equation (\ref{Lorentz_noncont}), 
we find that 
$(\Pi_0^*)^{-1} \equiv x_1 - x_2 P_*$
provides an estimate of $\Pi_0^{-1}$ in
the period range that includes the dip.
Fig \ref{fig:fitting_numerical} and Table \ref{tab:fitting_result} present the results of the fitting,
which are computed by the fit function of \textsc{gnuplot}.

We observe in Fig.~\ref{fig:fitting_numerical}
that the numerical results (blue filled circles)
are fitted by the Lorentzian functions (orange lines)
quite well except in the case of $X_c=0.60$,
in which the contribution of the wavy component
is significant in the numerical results.
In fact,
we find from Table \ref{tab:fitting_result} that
the relative uncertainties in $\tilde{\sigma}$ 
are 8 per cent for $X_c=0.60$ and less than 3 per cent for the other cases.

According to
Table \ref{tab:fitting_result},
$\tilde{\sigma}$ 
first
decreases rapidly 
along with the evolution
until $X_c=0.37$
(from $6.9\,\mathrm{h}$ for $X_c = 0.72$
to $1.2\,\mathrm{h}$ for $X_c = 0.37$). 
The initial decrease in $\tilde{\sigma}$ is
associated with growth of the convective core
(cf.~Fig.~\ref{fig:N2profile}).
Then, the decrease becomes
milder between $X_c=0.37$ and $X_c=0.18$.
Finally, $\tilde{\sigma}$ turns
to increase slightly
from $1.0\,\mathrm{h}$ for $X_c=0.18$ to 
$1.3\,\mathrm{h}$ for $X_c=0.06$.
The peak position
$P_*$ 
monotonically
gets smaller along with
the evolution,
though it changes less rapidly
than $\tilde{\sigma}$
between $3.0 \,\mathrm{d}$ for $X_c = 0.72$
and $2.1\,\mathrm{d}$ for $X_c = 0.37$. 
On the other hand,
the monotonic decrease in 
$\Pi_0^*$ is caused by the increase of
the maximum value of $N^2$, which is realised just outside the convective core.

\subsection{Evolution of $\tilde{\epsilon}$}
\label{subsec:four_one}

\begin{figure*}
    \centering
    \includegraphics[width=2\columnwidth]{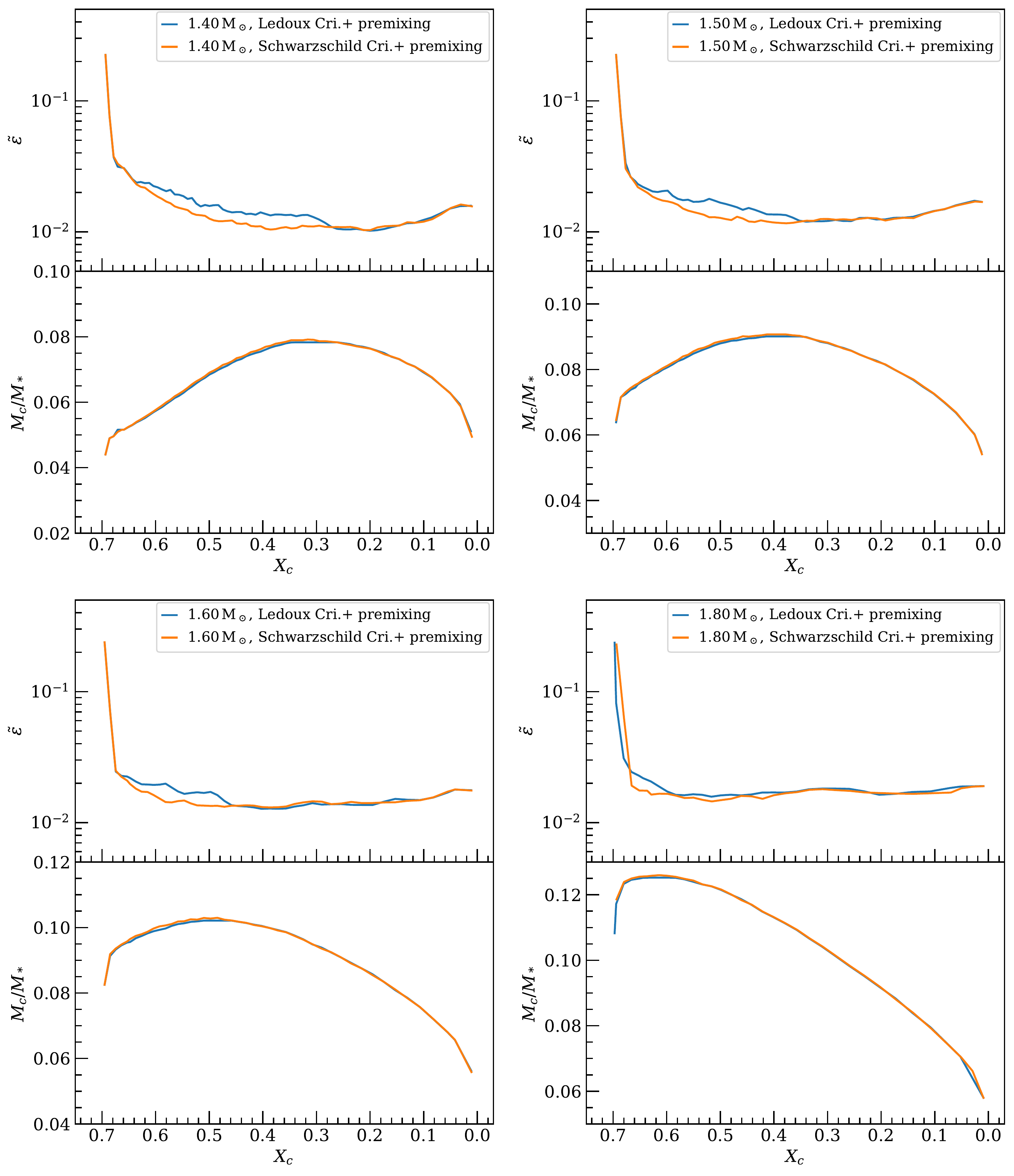}
    \caption{%
    Evolution
    of $\tilde{\epsilon}$ and $M_\mathrm{core}$ 
    as function of
    the central hydrogen mass fraction $X_c$ for the models with $1.4$ (upper left panels), $1.5$ (upper right panels), $1.6$ (lower left panels) and $1.8\,\mathrm{M}_{\odot}$
    (lower right panels). 
    The ordinates represent $\tilde{\epsilon}$ (upper panels) and $M_\mathrm{core}$ (lower panels). The core mass $M_\mathrm{core}$ is normalised by the total stellar mass $M_*$.
    Note that $X_c$ decreases 
    during the evolution from $0.70$
    at the zero-age main-sequence stage. 
    The blue and orange lines correspond
    to the Ledoux and Schwarzschild cases, respectively
    (refer to Appendix \ref{App:eps_tilde} for the details). We set the rotation period to $0.455 \, \mathrm{d}$ for all the cases.
    }
	\label{fig:Xcepsmcore}
\end{figure*}

In Section \ref{sec:three}, 
we have shown that $\epsilon$ (or $\tilde{\epsilon}$), which is defined as equation (\ref{epsilon}) (or equation \ref{tildeepsilon}), is the fundamental parameter that controls the strength of the interaction between the pure inertial waves in the convective core and the gravito-inertial waves in the radiative envelope.  Moreover,
it has a direct influence on
the width and depth of the dip structure
in the period-spacing.
We numerically
estimate this parameter in this section.
The details of the parameter settings 
of \textsc{mesa}
are given
in Appendix~\ref{App:eps_tilde}.

The top panels of Fig.~\ref{fig:Xcepsmcore} present
the evolution of $\tilde{\epsilon}$ 
during the main-sequence stage as functions of the central hydrogen mass fraction ($X_c$)
for the models with $1.4$, $1.5$, $1.6$ and $1.8 \,\mathrm{M}_{\odot}$, whereas the bottom panels show the corresponding evolution of the mass of the convective core ($M_{\mathrm{core}}$). 
The blue (orange) lines represent
the results obtained by
adopting
the Ledoux (Schwarzschild) criterion for convection
(cf.~Appendix~\ref{App:eps_tilde}).
Ignoring jaggy structures,
which probably reflect numerical errors, we
commonly observe for the 
three values of mass 
that, along with the evolution 
(from right to left in the figure), 
$\tilde{\epsilon}$ first decreases steeply in the early phase, takes its minimum at an intermediate stage, and then comes to increase very slowly in the later phase. On the other hand, $M_{\mathrm{core}}$ shows
the opposite behaviour.  It increases first and turns to decrease after taking its maximum near the stage when $\tilde{\epsilon}$ gets minimum.

The negative correlation between
$\tilde{\epsilon}$ and $M_{\mathrm{core}}$
can readily be understood.
We first note that the right-hand side limit $N_0$ is mostly determined by
the gradient of the mean molecular weight
in the mass coordinate.
As $M_{\mathrm{core}}$ 
increases in the early phase, 
the gradient becomes larger
near the inner boundary of the radiative envelope (cf.~Fig.~\ref{fig:N2profile}). When the core starts to shrink, the steep gradient that is
created during the growing phase
and left behind in the radiative core
is quickly smoothed out because of
the element diffusion.
In the later phase of core shrinkage,
the gradient of the mean molecular weight
on the radiative side of the boundary
gets milder
because the core shrinkage rate 
$|dM_{\mathrm{core}}/dX_c|$
increases.

Turning to the difference
between the 
Ledoux and Schwarzschild
cases,
we find 
in Fig.~\ref{fig:Xcepsmcore}
that,
for all of the masses,
$\tilde{\epsilon}$ is slightly larger
in the Ledoux
case than in the
Schwarzschild case
during the core growth
except for the very early stage,
though $M_{\mathrm{core}}$
has little difference between the two cases
for the all values of $X_c$.
As explained in \citet{paxton2019modules},
the convective premixing 
effectively reproduces
semiconvection,
which induces slow mixing 
in a shallow layer just outside the convective core
so that
the radiative temperature gradient is
equal to the adiabatic value.
While the both cases
take this process into account,
the resultant structure
depends on which convective criterion is adopted.%
\footnote{%
In the models we construct,
the boundary condition
that the radiative and adiabatic
temperature gradients
are equal to each other
on the convective side of
the boundary 
\citep[e.g.][]{2014A&A...569A..63G}
is more accurately satisfied 
in the Schwarzschild case than
in the Ledoux case.
}
The larger $\tilde{\epsilon}$ 
for the Ledoux case
is caused by
smaller $N^2$ in the layer,
which is realised by
smoother abundance profiles.
It is actually still a matter of debate
how we should formulate
 mixing processes at the boundary
between the convective core and
the radiative envelope.
While we do not intend to give the final answer
to this problem in this paper,
Fig.~\ref{fig:Xcepsmcore} clearly 
shows that the two different cases
can be discriminated by
measuring $\tilde{\epsilon}$
in the core growing phase of evolution%
. This particular example demonstrates
potential of $\tilde{\epsilon}$
as a diagnostic parameter in asteroseismology.

In the last column of
Table \ref{tab:fitting_result},
we present
$\tilde{\epsilon}$
of the models in Fig.~4 of
\citet{saio2021rotation}.
Comparing the evolutionary change
in $\tilde{\epsilon}$ with that in
$\tilde{\sigma}$,
we find that
the both parameters follow the similar trend.
They decrease rapidly from
$X_c = 0.72$ to $0.60$ 
by factors $3.0$ and $9.3$ for
$\tilde{\sigma}$ and $\tilde{\epsilon}$,
respectively,
and then
mildly from $X_c = 0.60$ to $0.37$
with nearly the same factor of $\sim 2$.
Then, they show little variation
between $X_c = 0.37$ and $0.18$
with $-10$ and $+3$ per cent decrease
in $\tilde{\sigma}$ and $\tilde{\epsilon}$,
respectively. Finally, they both show increase between $X_c = 0.18$ and $0.06$
by $+21$ and $+11$ per cent
in $\tilde{\sigma}$ and $\tilde{\epsilon}$,
respectively.  
We therefore understand that,
except for the initial phase when the convective core grows rapidly (cf.\ Fig.~\ref{fig:Xcepsmcore}), the change in $\tilde{\sigma}$ is mainly controlled by that in $\tilde{\epsilon}$, which implies
that $\tilde{V}$ does not change significantly
(cf.\ equation~\ref{sigmatilde}).

\section{Discussion}
\label{sec:five}

\subsection{Asteroseismology of the dip}
\label{subsec:five_one}

We have analysed how the dip structure
in the $P$--$\Delta P$ diagram is formed
as a result of the interaction
between pure inertial waves in the convective core
and gravito-inertial waves in the radiative envelope
(forward problem).
Once we understand the physics of the dip formation,
we may ask what kind of structure information we can derive
from the dip (inverse problem).
We find that the corresponding hump structure
in the $P$--$1/\Delta P$ diagram
is approximately described by
a normalised Lorentzian function. 
While we consider two cases,
those of continuous and discontinuous profiles,
they agree in the
functional form of the hump structure, which
is given by
equations
(\ref{Lorentz}) and (\ref{Lorentz_noncont}),
respectively.
For simplicity, we base our discussion
in this section
on equation
(\ref{Lorentz_noncont})
for the case of discontinuous profiles.

Because
$\tilde{\sigma}$ is essentially the ratio
between 
$\tilde{\epsilon}/\tilde{V}$ and $\Omega$ (the rotation rate of the radiative envelope) (cf.~equations \ref{Vtilde_def} and \ref{sigmatilde}),
and
$\Omega$
can be estimated independently from the frequency spectrum outside the dip,
we may regard
$P_*$ (the period of the pure inertial mode in the core) 
and
$\tilde{\epsilon}/\tilde{V}$ 
as two structural parameters that we can derive
from the dip (or hump). 
Strictly speaking,
we cannot determine $\tilde{\epsilon}$ and $\tilde{V}$ separately
without any additional assumptions.
In fact,
although
function $\tilde{F}$, on which $\tilde{V}$ depends (cf.~equation \ref{Vtilde_def}),
is explicitly given by
equation (\ref{F_uniformdensity})
in the case of uniform-density cores,
it generally depends
on the core structure
through $g_{\ell}$ and $h_{\ell}$ (cf.~equation \ref{tildeF}).
However, since
the evolutionary change in
$\tilde{\sigma}$
largely depends on
that in $\tilde{\epsilon}$
(cf.\ discussion at the end of
Section \ref{subsec:four_one}),
we may estimate
the relative change
in the Brunt--V\"ais\"al\"a frequency
in the boundary layers
from that in the dip structure.

As we have discussed in Section \ref{subsec:four_one},
since we do not understand
the mixing process 
in stars completely,
there exist
uncertainties in the estimate of
$\tilde{\epsilon}$ in the forward problem based on the evolutionary models.
On the other hand, if the observed frequencies are precise enough, we may measure the shape of the dips so accurately
that we can estimate $\tilde{\epsilon}/\tilde{V}$ reliably.
In this sense, the inverse problem could be more useful than the forward problem
to probe the structure at the boundary between the convective core and the radiative envelope of $\gamma$ Dor stars.

\subsection{Roles of the geometric property}
\label{subsec:five_two}

\citet{ouazzani2020first} suggest that
the resonance between a pure inertial mode in the core
and a gravito-inertial mode in the envelope
could mainly be controlled by
the geometric factor,
which measures how close their angular structures are to each other.
In the case of the uniform-density core,
this factor is simply given 
in our formulation
by $c_{k, \ell}$ that is defined by equation (\ref{def_ckl})
(provided that the Hough function $\Theta_k^m$ is properly normalised).
\citet{ouazzani2020first} regard that
the interaction between the core and envelope modes
with larger values of $c_{k,\ell}$
leads to stronger resonance, which
corresponds to wider and shallower dips
in the $P$--$\Delta P$ diagram.
On the other hand,
the analysis in the present paper
demonstrates that
the width (as well as the depth)
of the dips is essentially fixed by
$\tilde \sigma$, which is determined by
$\tilde{\epsilon}/ \tilde{V}$ and $\Omega$ (the rotation rate of the envelope).
We actually consider $\tilde{\epsilon}$ as the most important parameter
of the interaction,
though it is not related to the geometrical factor, but is
essentially controlled
by the chemical composition profiles
near the boundary between the core and the envelope.
Still, we admit that
the geometric factor is also important in our formulation
in two respects.
The first point is found
when we approximate
the matching condition of
equation (\ref{detMN})
by
equation (\ref{zeropoint}).
In this approximation,
we replace the determinant of an infinite matrix
by the most dominant element.
For this to be valid,
we have to assume that
the geometry factor, $c_{k, \ell}$, 
which is a common factor of the terms in the most dominant element,
is not negligibly small.
On the other hand,
we observe the second point in equation (\ref{tildeF}).
The numerator and denominator of
the right-hand side of this equation
are essentially equal to
the integrals of $\xi_{r,\mathrm{con}} \Theta_k^m$ 
and $p'_{\mathrm{con}} \Theta_k^m$, respectively,
over the sphere at $r=R_{\mathrm{core}}$.
They therefore provide measures
of how close the angular structures
of $\xi_{r,\mathrm{con}}$ and $p'_{\mathrm{con}}$
are to $\Theta_k^m$
at the outer boundary of the convective core.
Since $V$ depends on the ratio of these measures
(cf.~equation \ref{V_def}),
the width and the depth of the dip structure,
which are controlled by 
$\tilde{\epsilon}/\tilde{V}$%,
cannot simply be determined by 
the angular structure of 
either $\xi_{r,\mathrm{con}}$ or $p'_{\mathrm{con}}$.

\subsection{Impact of the convective overshooting}
\label{subsec:five_three}

\citet{saio2021rotation} examined how
the convective overshooting influences
the dip structure in the 
$\nu_{\mathrm{co-rot}}$--$\Delta P_{\mathrm{co-rot}}$
diagram.
Taking the diffusive overshooting 
\citep{herwig2000evolution}
into account,
they showed that a single (isolated) dip structure
in the case of no convective overshooting
turns into multiple dips.
We reconsider here 
this phenomenon based on the picture 
presented in this paper.
More specifically,
we rely on the analysis for continuous profiles
in Section \ref{subsec:three_three}
because the overshooting causes additional
mixing.
As \citet{saio2021rotation} demonstrate
in their Figs.~8 and 9,
the diffusive convective overshooting
generates
a thin intermediate radiative layer
between the convective core
and
the steep chemical composition gradient layer in the radiative envelope.
There thus exist three separate regions of wave propagation.
However,
because the constant period-spacing
(outside the dip structure) comes from
the propagation in
the radiative region above the steep chemical composition gradient,
we may treat 
the inner two regions
(the convective core and
the intermediate radiative zone)
together
to discuss the formation of the dip structure.
In order to understand the relation between the dips
and the wave propagation,
we pay attention to the case of
$M=1.50\,\mathrm{M}_{\odot}$,
$f_{\mathrm{ov}}=0.02$%
\footnote{%
\citet{saio2021rotation} use
 $h_{\mathrm{os}}$ for $f_{\mathrm{ov}}$.
}
(the larger overshooting parameter)
and $X_{\mathrm{c}}=0.06$,
in which three dips appear 
around $0.23\,\mathrm{d}^{-1}$, 
$0.31\,\mathrm{d}^{-1}$ and $0.40\,\mathrm{d}^{-1}$
in
the $\nu_{\mathrm{co-rot}}$--$\Delta P_{\mathrm{co-rot}}$ diagram
\citep[cf.\ green filled circles in the upper right panel of Fig.~7 of][]{saio2021rotation}
.
Observing
the profiles of the corresponding radial displacement $\xi_r$
around the centre of each dip
in the right panel of their Fig.~9,
we find 
that
the number of nodes
increases 
one by one
in the inner two layers
as $\nu_{\mathrm{co-rot}}$ decreases,
though they are located
only in the intermediate radiative layer
and the upper part of the convective core.
Based on this observation,
we speculate that,
due to the appearance of
the intermediate radiative layer,
where gravito-inertial waves can propagate,
a pure inertial mode in the convective core
is split into multiple eigenmodes
in the inner two regions,
and that
each of the split modes interacts
with the gravito-inertial modes
in the outer radiative region
above the steep chemical composition gradient
to produce a dip
in the $P$--$\Delta P$ (or $\nu_{\mathrm{co-rot}}$--$\Delta P$)
diagram
based on the mechanism that
we propose in Section \ref{sec:two}.
We therefore expect to find more dips 
(with more nodes of the corresponding $\xi_r$)
in the diagram
if we extend the diagram to the lower-frequency
(longer-period) range
than are shown in \citet{saio2021rotation}.
As they discuss,
if the overshooting does not create
the intermediate radiative layer
but just extends the convective core
as in the case of 
convective penetration \citep{zahn2002convective},
the number of dips
in the $P$--$\Delta P$ diagram
does not change.
Therefore,
we may get information about the overshooting process
by 
checking
whether the dip is split or not.
\citet{saio2021rotation}
also point out that
the width and depth of dips in the case of overshooting
are broader and shallower, respectively,
which implies that the interaction gets stronger,
than in the case without overshooting.
In this case, we may consider
the interaction of waves
between the two inner regions
(the convective core and the intermediate
radiative region).
Because
the shape of the dips are controlled by
$\epsilon/V$
(see Section \ref{sec:three}),
we may interpret that
the shape change is because of the increase in $\epsilon$,
which is in turn caused by
the 
shallower gradient of $N^2$
in the intermediate radiative region.
Figs.~8 and 9 
of \citet{saio2021rotation}
are consistent with this interpretation.

\subsection{Frequency dependence of period-spacing}
\label{subsec:five_four}

In Section \ref{sec:three}, we assumed
that the period-spacing of the gravito-inertial mode $\Pi_0$ is constant 
irrespective of
period $P$ (or the spin parameter $s$) for simplicity. However,
the numerical results given in Fig.~4 of \citet{saio2021rotation} 
suggest that $1/\Delta P$ (or $\Delta P$)
generally decreases (or increases) with $P$.
To take this property into account, we adopted linear function $x_1-x_2 P$ for fitting the base line of the hump
structure in
$1/\Delta P$ in Section \ref{subsec:four_two}. 
We discuss here the cause of this tendency. 
It should be noted that, in this section, $\Delta P$ represents the period-spacing with the frequency dependence,
and that
$\Pi_0$ means the constant period-spacing
in the limit of an infinite period
(cf.~equation \ref{Pi_0}).

We can give three
possible
ideas for explaining the 
frequency dependence
of $\Delta P$.
The first one
is lack of resolution in the numerical calculations,
which results in the wavelengths of oscillations comparable to (or even smaller than) the mesh sizes of equilibrium models. \citet{saio2021rotation} point out that this is the reason for the increase in
the period-spacings particularly in the low frequency
(large $s$) limit, for which the 
radial wavelengths are extremely short.
The second idea
is low accuracy of the asymptotic formula of $\Pi_0$ given by equation (\ref{Pi_0}).
In fact, we demonstrate in Appendix~\ref{App:A}
that the formula can be corrected to explain
the non-constant period-spacings.
The third idea is the wavy component caused by rapid change in the equilibrium structure
(see Fig.~\ref{fig:fitting_numerical}, especially the case of $X_c=0.60$). 
This component is not taken into account in the asymptotic analysis in Appendix~\ref{App:A}.

In order to examine which of the three reasons are more important in the frequency range in which the dip structure appears, we perform two additional analyses.
We first compute the period-spacings under TAR
for the same evolutionary models constructed by \citet{saio2021rotation}, but with increased
spatial
resolution by interpolation.
We call $\Delta P$ obtained by
this analysis as $\Delta P_{\text{TAR IR}}$.
We secondly calculate the frequencies
by solving the following asymptotic condition directly:
\begin{equation}
    \int_{r_a}^{r_b} k_r dr = (n + \hat{\alpha}) \pi
    ,
    \label{eq:asymptotic_eigenvalue_condtion}
\end{equation}
in which we assume $\hat{\alpha} = 0$ for simplicity.
We refer to $\Delta P$ computed by this method as $\Delta P_{\text{ASYMP}}$.
We note that equation (\ref{Pi_0}) can be derived from equation (\ref{eq:asymptotic_eigenvalue_condtion})
for low frequencies.

We confirm
in the frequency range between
$0.2$ and $1\,\mathrm{d}^{-1}$
that 
$\Delta P$ by \citet{saio2021rotation}
generally agrees with
$\Delta P_{\text{TAR IR}}$
in the high frequency range,
whereas
the former gets larger than the latter 
as the frequency becomes lower.
This difference, which is caused by
the numerical errors,
tends to increase for smaller $X_c$.
In fact,
it is about $0.3$ ($5$) per cent for $X_c=0.72$ ($0.18$)
at $0.2\,\mathrm{d}^{-1}$.
We also find 
in the same frequency range
that
$\Delta P_{\text{ASYMP}}$
can be reproduced by
the corrected asymptotic formula
of equation (\ref{Pi_0_morecomp}) in Appendix~\ref{App:A}
nearly completely.
In addition,
we understand
that
$\Delta P_{\text{ASYMP}}$
almost converges on
$\Delta P_{\text{TAR IR}}$
in the low frequency limit as expected,
and that
both of them slightly decrease
for higher frequencies.
Their discrepancy in the high frequency range
can be interpreted as
the contributions of the wavy component
and 
the inaccuracy of the asymptotic description.

Based on these considerations,
we draw the following conclusions:
the slopes of the baseline
for $X_c = 0.72$ and $0.60$ in
Fig.~\ref{fig:fitting_numerical}
can mostly be explained by
the corrected asymptotic formula of 
$\Delta P$ and the wavy component,
whereas
those for 
$X_c = 0.37$ and $0.18$
have non-negligible influences
of the resolution problem.
However, in any case,
the variation in the period-spacing is small enough
in the frequency range of each dip to be approximated by a linear function, as we assume in Section \ref{subsec:four_two}.

One message in this section is that the approximation 
of constant period-spacing,
which is widely assumed in the analysis of $\gamma$ Dor stars, may not be accurate enough.
This is particularly the case
when we analyse the observed frequencies
with very high precision,
as those provided by \Kepler.
The corrected form of $\Delta P$ due to higher-order terms should be considered in such cases (cf.~equation \ref{Pi_0_morecomp} or \ref{Pi_0_comp} in
Appendix~\ref{App:A}).

\subsection{Comparison with mixed modes in evolved stars }
\label{subsec:five_five}

We may compare
the resonantly coupled modes in $\gamma$ Dor stars,
which are analysed in the present paper,
with
mixed modes in subgiant and red giant stars
\citep[e.g.][]{mosser2012,hekker2018,appourchaux2020}.
In fact, the both types of modes are constructed
by the interaction between two 
kinds of waves 
with different physical characters, which are trapped in different regions of stars.
In the case of $\gamma$ Dor stars,
the two kinds of waves are pure inertial waves in the convective core
and gravito-inertial waves in the radiative envelope.
On the other hand,
the mixed modes in the evolved stars are composed of
gravity waves in the core
and
acoustic waves in the envelope.

The eigenmode condition for the mixed modes is formally given by
\begin{equation}
\tan \Theta_{\mathrm{p}}
\cot \Theta_{\mathrm{g}}
=
q
,
\label{mixed_mode_condition}
\end{equation}
in which $\Theta_{\mathrm{p}}$ and $\Theta_{\mathrm{g}}$
are phase functions of acoustic and gravity waves, respectively.
Following the dispersion relations of the waves,
$\Theta_{\mathrm{p}}$ and $\Theta_{\mathrm{g}}$ are
linear functions of the frequency and the period of oscillations,
respectively.
Note that equation (\ref{mixed_mode_condition})
is derived in the asymptotic limit
by assuming that the wavelengths of the constituent waves
are much shorter than the scale height
of the equilibrium structure.
This assumption is accurate for red giants
and in the late stage of subgiants.
The left-hand side of equation (\ref{mixed_mode_condition})
is a product of
the core variable $\cot\Theta_{\mathrm{g}}$
and
the envelope variable $\tan\Theta_{\mathrm{p}}$,
whereas
the right-hand side
is the coupling factor $q$,
which is determined by
the properties of the intermediate evanescent region \citep[cf.][]{unno1989nonradial,2016PASJ...68..109T,2020A&A...634A..68P}.
In fact,
we observe
the same structure in
equation (\ref{pict}).
The first part of the left-hand side of this equation,
$-V (s - s_*)$,
depends on the properties of the pure inertial waves
in the convective core,
while
the second part in the square bracket
represents the physical character of the
gravito-inertial waves in the radiative envelope.
In addition,
$\epsilon$ on the right-hand side
is the control parameter of the interaction,
which is determined by the structure of the interface
between the convective core and the radiative envelope.
We may therefore regard
the resonantly coupled modes in
$\gamma$ Dor stars
as another kind of mixed modes.

\section{Conclusion}
\label{sec:six}

We have discussed the formation process of
the dip structure
in the period and period-difference diagram
of $\gamma$ Dor stars,
which is recently found in
the observations
 \citep{saio2018astrophysical}
and
is reproduced in the numerical calculations
 \citep{ouazzani2020first,saio2021rotation}.
We have first presented a simple and qualitative picture
of the dip formation
that relies on the following two key processes:
(1)
insertion of a pure inertial mode of the convective core
into the frequency spectrum of gravito-inertial modes of the
radiative envelope, which exhibits equal spacing in period,
and
(2)
relaxation of the modes in the spectrum
due to the avoided crossings.
Based on the picture that
the dip is generated as a result of interaction
between the pure inertial waves in the convective core
and the gravito-inertial waves in the radiative envelope,
we have formulated the problem by using approximate analytic expressions
of the wave solutions in each region.
We have found that
the main control parameter of the interaction
is $\epsilon$ 
(or $\tilde{\epsilon}$
if the structure is discontinuous)
that is 
inversely 
correlated with the variation in
the squared 
Brunt--V\"ais\"al\"a frequency
at the convective-core boundary.
The shape of the dip
can approximately be described by
the Lorentzian function whose width and height depend on 
$\epsilon$
(or $\tilde{\epsilon}$)
in the sense that 
the width and height are broader and lower, respectively,
as $\epsilon$ 
(or $\tilde{\epsilon}$)
is larger.
We have also
demonstrated based on the evolutionary calculations
that
the change of 
$\tilde{\epsilon}$
is almost synchronised with
the mass of the convective core.
In particular,
$\epsilon$ becomes minimum, which means that the interaction becomes minimum,
when the mass of the convective core becomes maximum
during the main-sequence evolution.
Because the stellar structure
just outside the convective core 
sensitively depends on
various physical processes
including
diffusion,
rotational mixing
and
convective overshooting,
the information that we obtain from the observed dip
structure
would provide precious constraints on these processes,
which still contain some theoretical uncertainties.

\section*{Acknowledgements}
We thank Hideyuki Saio for providing the numerical data and constructive
comments that helped a lot to improve this paper. 
We also acknowledge Umin Lee 
and Francois Ligni\`eres
for their insightful comments about this study.
TT is supported by IGPEES, WINGS Program, the University of Tokyo.
MT is grateful for
the financial support by
JSPS KAKENHI Grant Numbers JP18K03695
and JP22K03672.

\section*{Data availability}
The data underlying this article will be shared on reasonable request to the corresponding author.

%%%%%%%%%%%%%%%%%%%%%%%%%%%%%%%%%%%%%%%%%%%%%%%%%%



\bibliographystyle{mnras}
\bibliography{reference} % if your bibtex file is called example.bib

%%%%%%%%%%%%%%%%%%%%%%%%%%%%%%%%%%%%%%%%%%%%%%%%%%

%%%%%%%%%%%%%%%%% APPENDICES %%%%%%%%%%%%%%%%%%%%%

\appendix

\section{Derivation of non-constant period-spacing}
\label{App:A}

The period-spacing of Kelvin g modes
with a given azimuthal order
is almost constant
in the co-rotating frame
in the asymptotic limit
of a large spin parameter,
or equivalently a low frequency
(cf.\ equation \ref{Pi_0}).
The deviation from this limit
is, however, not negligible
when we 
analyse the dip structure
as in section \ref{subsec:four_two}.
We therefore discuss
in this section
the frequency dependence of
the period-spacing 
based on the asymptotic analysis
in the traditional approximation of rotation.

\begin{figure}
	\includegraphics[width=\columnwidth]{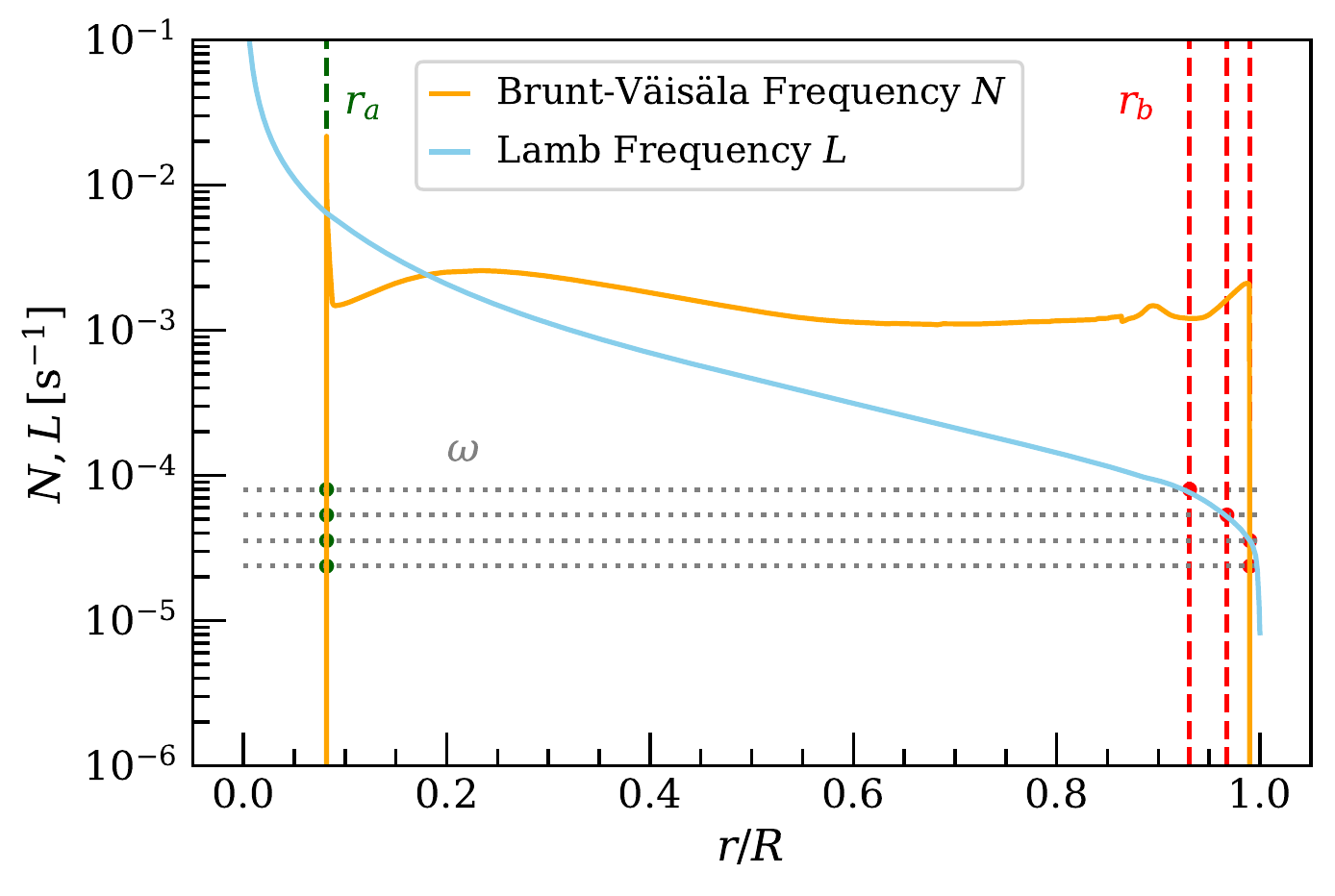}
    \caption{%
    Typical propagation diagram
    of $\gamma$ Dor stars
    in the co-rotating frame.
    Profiles of Brunt-V\"{a}is\"{a}la frequency $N$ (orange solid line) and Lamb frequency $L$ (blue solid line) are calculated by \textsc{mesa}. Also indicated are the locations of $r_a$ (vertical green dashed lines) and $r_b$ (vertical red dashed lines) for each frequency $\omega$ (horizontal gray dotted lines). 
    Horizontal and vertical axes
    represent the
    distance from centre (normalised by
    the stellar radius) and the frequency, respectively.Note that we set the stellar mass $M_* = 1.5 \, \mathrm{M}_\odot$ , the central hydrogen mass fraction $X_c = 0.58$ and the rotation period $2\pi/\Omega = 0.455 \, \mathrm{d}$.
    }
    \label{fig:rf}
\end{figure}

\begin{figure*}
    \centering
	\includegraphics[width=2\columnwidth]{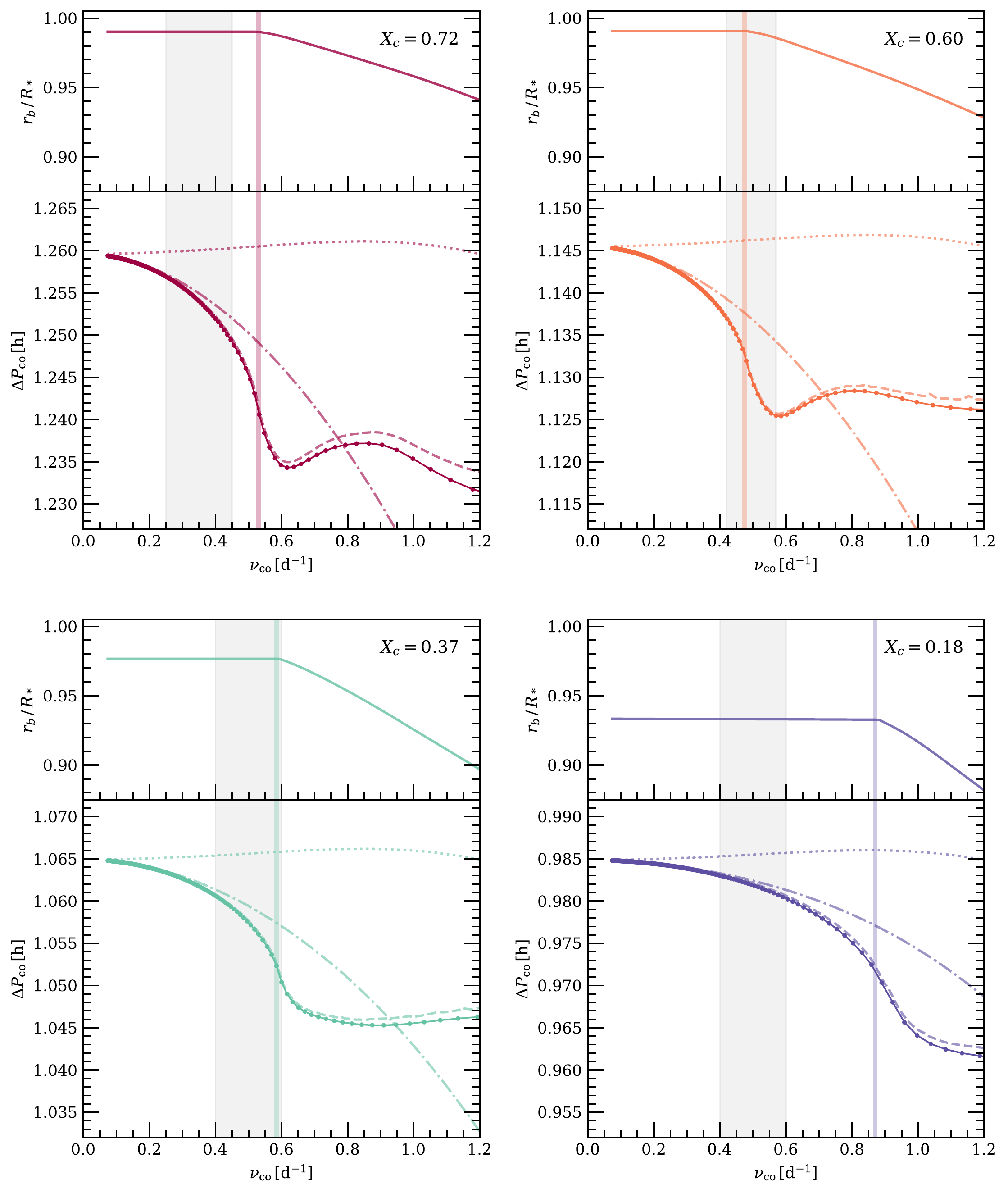}
    \caption{
    Frequency dependence of the
    period-spacing
    ($\Delta P$) in the co-rotating frame
    and the outer boundary of the propagative cavity
    ($r_b$)
    of Kelvin g modes with $m=1$
    for the four
    evolutionary models 
    in Fig.~4 of \citet{saio2021rotation}
    with the mass of $1.5\,\mathrm{M}_{\odot}$,
    the rotation period of $0.455 \, \mathrm{d}$ and the central hydrogen mass fractions of $X_c = 0.72$ (ZAMS), $0.60$, $0.37$ and $0.18$.(%
    upper part of each panel%
    ) 
    $r_b$ in the unit of the total stellar radius $R_*$ (solid lines) as a function of
    the frequency $\nu$.
    The vertical line represents
    the boundary 
    below and above which
    $r_b$ is
    determined by $\omega=N$ and by $\omega=L$, respectively.(lower part of each panel)
    $\nu$--$\Delta P$ diagram obtained by the numerical results based on equation (\ref{eq:asymptotic_eigenvalue_condtion})
    (the filled circles
    connected by solid lines)%
    .
    Note that $\nu$ means
    the reciprocal of the average period of two neighbouring solutions $[(P_{n+1}+P_n)/2]^{-1}$. The dashed, 
    dashed-dotted and dotted 
    curves are drawn based on
    equations  (\ref{Pi_0_morecomp}), (\ref{Pi_0_comp})
    and (\ref{Pi_0_simp}), respectively.
    We fix $r_b$ at the base of the convective envelope in equation (\ref{Pi_0_comp}).
    The grey shaded zone corresponds to
    the area where the dip 
    structure is located.}
    \label{fig:nurbdp}
\end{figure*}

\begin{figure}
    \centering%
	\includegraphics[width=\columnwidth]{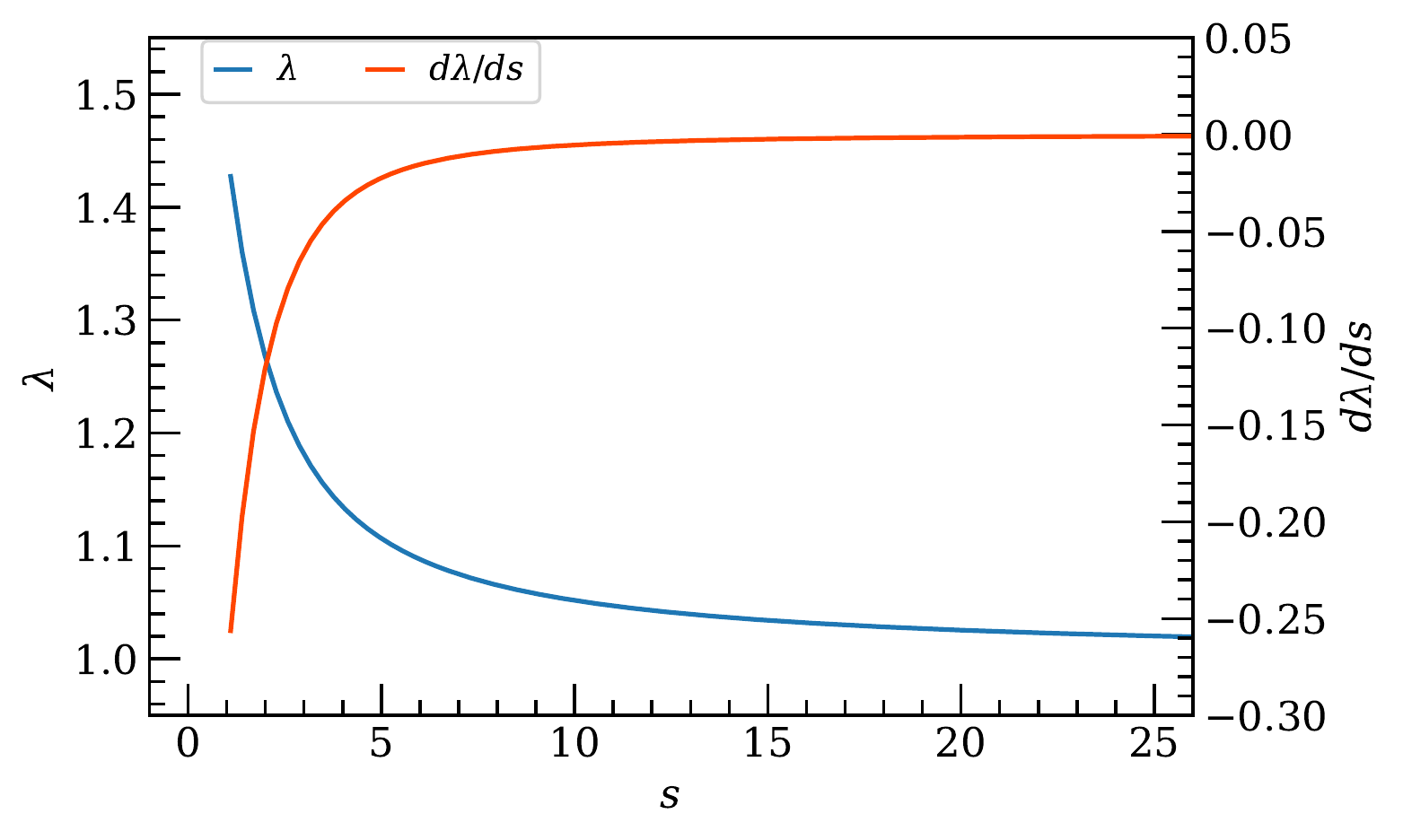}
    \caption{%
    Eigenvalue of the Laplace tidal equation
    $\lambda$ (blue line) and 
    its derivative
    $d \lambda / ds$ (orange line) 
    as a function of
    the spin parameter $s$ in the case of Kelvin
    g modes
    with the azimuthal order of $1$. }
    \label{fig:lamdlamds}
\end{figure}

We first note
the frequency dependence of
two parameters $r_b$ and $\lambda$ in
equation (\ref{Pi_0}).
In Fig.~\ref{fig:rf},
we depict how
the inner and outer boundaries, $r_a$ and $r_b$,
of the propagative cavity
are determined
by the profiles of
the Brunt--V\"ais\"al\"a frequency, $N$,
and
the Lamb frequency, $L=\sqrt{\lambda}c_s/r$,
for a given frequency $\omega$
in a typical model of $\gamma$ Dor stars.
We observe that $r_a$ is fixed by $N$ and almost
equal to the top of the convective core
independently of the frequency,
whereas $r_b$ is constrained by the condition of
$\omega = \min\left(N, L\right)$.
As we see in the top panel of Fig.~\ref{fig:nurbdp},
$r_b$ is nearly constant in the low frequency range,
because the condition $\omega=N$ 
essentially sets $r_b$ to the base of
the convective envelope.
In contrast,
$r_b$ decreases almost linearly
as a function of the frequency
in the high frequency range,
where $r_b$ is fixed by $\omega=L$.
On the other hand,
Fig.~\ref{fig:lamdlamds} shows
how the eigenvalue of the Laplace tidal equation,
$\lambda$, and its derivative depend on the spin parameter
$s$.
We observe that $\lambda$ is only a weak function of $s$  for $s \gtrsim 10$.

We next analyse
equation (\ref{eq:asymptotic_eigenvalue_condtion})
to derive the expression of the period-spacing.
We find from equation (\ref{kr})
that $k_r$ can be regarded as a function of
$\omega$, $\lambda$ and $r$.
We can therefore introduce
\begin{equation}
    K\left(r_b, \omega, \lambda\right)
    \equiv
    \int_{r_a}^{r_b}
    k_r\left(\omega, \lambda; r\right)
    dr
    .
\end{equation}
If we express the angular
frequency of the mode
with radial order $n$ as $\omega_n$
and the corresponding $\lambda$ as $\lambda_n$,
we obtain from equation (\ref{eq:asymptotic_eigenvalue_condtion})
\begin{equation}
    K\left(r_b(\omega_{n+1}), \omega_{n+1}, \lambda_{n+1}
    \right)
    -
    K\left(r_b(\omega_{n}), \omega_{n}, \lambda_{n}
    \right)
    = 
    \pi
    ,
    \label{eq:dkr_int}
\end{equation}
where we have assumed that
$r_a$ and $\hat{\alpha}$ are independent of the frequency.
Under the assumption that
the frequency difference,
$|\omega_{n}-\omega_{n+1}|$, is small,
equation (\ref{eq:dkr_int}) can be approximated by
\begin{equation}
\frac{\partial K}{\partial r_b}
\Delta r_b
+
\frac{\partial K}{\partial \omega}
\Delta\omega
+
\frac{\partial K}{\partial \lambda}
\Delta\lambda
=
\pi
,
\label{eq:pdK}
\end{equation}
in which 
we have defined
\begin{align}
\Delta r_b & = r_b(\omega_{n+1}) - r_b(\omega_{n})
,
\\
\Delta\omega & = \omega_{n+1} - \omega_{n}
\end{align}
and
\begin{align}
\Delta\lambda & = \lambda_{n+1} - \lambda_{n}
.
\end{align}
Note that
all of the
partial derivatives in
equation (\ref{eq:pdK})
should be evaluated
for
\begin{align}
r_b & = 
[r_b(\omega_{n})+r_b(\omega_{n+1})]/2,
\\
\omega & = 
(\omega_{n} + \omega_{n+1})/2
\end{align}
and
\begin{align}
\lambda & =
(\lambda_{n}+\lambda_{n+1})/2
.
\end{align}
Using the two relations,
\begin{equation}
    \frac{\partial K}{\partial r_b}
    =
    k_r\left(\omega, \lambda; r_b\right) =
    0
    ,
\end{equation}
which comes from the definition of $r_b$,
and
\begin{equation}
\Delta\lambda
= 
\frac{d\lambda}{ds}
\Delta s
=
-
s \frac{d\lambda}{ds}
\frac{\Delta\omega}{\omega}
,    
\end{equation}
where $\Delta s = s(\omega_{n+1}) - s(\omega_{n})$,
we obtain from equation (\ref{eq:pdK})
\begin{equation}
\Delta P
=
\frac{2\pi^2}{{\omega}}
\left[
\int_{r_a}^{r_b}
\left(
s \frac{d\lambda}{ds}
\frac{\partial k_r}{\partial \lambda}
- {\omega}\frac{\partial k_r}{\partial \omega}
\right)
dr
\right]^{-1}
,
\label{eq:Pi0_raw}
\end{equation}
where
$\Delta P \equiv 2 \pi/\omega_{n+1}-2\pi/\omega_{n}$. Assuming $\omega \ll N$, which is valid in the propagative region of the high-order gravito-inertial modes
(except very close to the turning points),
we may approximate $k_r$ given by equation (\ref{kr}) as
\begin{gather}
    k_r \simeq \frac{ \sqrt{\lambda} N}{r \omega } \sqrt{ 1- \frac{r^2 \omega^2}{ \lambda c_s^2} }. \label{kr_more}
\end{gather}
Substituting equation 
(\ref{kr_more}) into equation (\ref{eq:Pi0_raw}),
we obtain 
\begin{equation}
    \Delta P
    \simeq
    2\pi^2
    \left[
    \sqrt{\lambda}
    \left( 
    1 + \frac{s}{2\lambda}\frac{d\lambda}{ds}
    \right)
    \int_{r_a}^{r_b}
    \frac{N}{r}
    \left(
    1 - \frac{r^2\omega^2}{\lambda c_s^2}
    \right)^{-\frac{1}{2}}
    dr
    \right]^{-1}
    .
    \label{Pi_0_morecomp}
\end{equation}
Although
this expression is quite accurate
for a wide range of the frequency
(cf.~Fig.~\ref{fig:nurbdp}),
the frequency dependence is not separated
from the integral of structure variables.
We may then assume
$r^2 \omega^2 / \lambda c_s^2 \ll 1$
to obtain
\begin{equation}
    \Delta P
    \simeq 
    2\pi^2\left( \int_{r_a}^{r_b} N \, \frac{dr}{r} \right)^{-1} \left[ \sqrt{\lambda}
    \left( 
    1 + \frac{s}{2\lambda} \frac{d \lambda }{ds} 
    \right)
    \left( 1+ \frac{\omega^2 \delta}{2\lambda} \right)
    \right]^{-1}, \label{Pi_0_comp}
\end{equation}
where $\delta$ is defined by
\begin{equation}
    \delta
\equiv \left( \int_{r_a}^{r_b} \frac{r N}{c_s^2} \, dr \right) \left( \int_{r_a}^{r_b} N \, \frac{dr}{r} \right)^{-1}. \label{delta2}
\end{equation}

If we set $\delta = 0$ in equation (\ref{Pi_0_comp}), we obtain
\begin{gather}
    \Delta P 
    \simeq 2\pi^2 \left( \int_{r_a}^{r_b} N \, \frac{dr}{r} \right)^{-1} \left[ \sqrt{\lambda}  \left( 1 + \frac{s}{2\lambda} \frac{d \lambda }{ds} \right) \right]^{-1}, \label{Pi_0_simp}
\end{gather}
which is the same expression as equation (10) of \citet{Ballot2012}. 
Considering the approximated expression 
of $\lambda \simeq m^2[1 + (2ms-1)^{-1}]$ \citep{townsend2003asymptotic}, we can confirm that the right-hand side of equation (\ref{Pi_0_simp}) is nearly constant for large $s$ (see also Fig.~\ref{fig:nurbdp}). It is therefore essential to consider 
$\delta$ for the frequency variation of the period-spacing.

In Fig.~\ref{fig:nurbdp}, we compare the numerical solutions of equation 
(\ref{eq:asymptotic_eigenvalue_condtion}) 
with the expressions by equations (\ref{Pi_0_morecomp}), (\ref{Pi_0_comp}) and (\ref{Pi_0_simp})
for 
Kelvin g modes with $m=1$ of
the evolutionary models constructed
by \citet{saio2021rotation}. 
We first observe that,
at each evolutionary stage
(specified by $X_c$),
the numerical solution and
all of the expressions
converge
on the same value
in the low frequency limit.
We actually obtain
from 
equations (\ref{Pi_0_morecomp}), (\ref{Pi_0_comp}) and (\ref{Pi_0_simp})
\begin{gather}
\lim_{s \to \infty}  \Delta P =
\frac{2\pi^2}{m} \left( \int_{r_a}^{r_b} N \, \frac{dr}{r} \right)^{-1} 
\label{eq:dP_limit}
\end{gather} 
in which 
$r_b$ essentially
means the radius of the
base of the near-surface convective zone.
The limiting value given by
equation (\ref{eq:dP_limit}) corresponds
to $\Pi_0$ introduced by
equation (\ref{Pi_0}).
We 
next
find that equation (\ref{Pi_0_simp}) 
poorly describes the frequency dependence 
of $\Delta P$
because it
provides almost constant values
(dotted curves). 
On the other hand, we notice that equation (\ref{Pi_0_morecomp}) reproduces the numerical solutions within $0.2$ per cent  in the entire frequency range
(dashed curves),
and also that equation (\ref{Pi_0_comp}) with constant $r_b$ (fixed at the base of the convective envelope) provides good approximations to the numerical solutions only in the low frequency range
(dashed-dotted curves).
Particularly
in the frequency ranges where
the dip structure is found
(the shaded areas in Fig.~\ref{fig:nurbdp}),
equation (\ref{Pi_0_comp}) differs from the numerical solutions by $0.8$ per cent at most.
The discrepancy
between the results by equations (\ref{Pi_0_morecomp}) and (\ref{Pi_0_comp})  (dashed and dashed-dotted curves, respectively,
in Fig.~\ref{fig:nurbdp})
can be explained by two reasons.
One is 
that the approximation of $r^2 \omega^2 / \lambda c_s^2 \ll 1$, which is used to derive equation (\ref{Pi_0_comp}),
becomes inaccurate as the frequency is higher.
The other is that $r_b$, which is assumed to be constant in equation (\ref{Pi_0_comp}),
depends on frequency
in the high frequency range
where $r_b$ is fixed by $\omega = L$.
The lower bound of this frequency region
is indicated by the vertical line
in each panel of Fig.~\ref{fig:nurbdp}.
This frequency dependence of $r_b$
is the reason for
the non-monotonic variations in $\Delta P$
predicted by equation (\ref{Pi_0_morecomp}).
We note that
the first effect
(the inaccuracy of
$r^2 \omega^2/(\lambda c^2_s) \ll 1$)
starts to appear in the lower frequency range
than the second effect
(the frequency dependence of $r_b$)
for all the values of $X_c$.

\section{Calculation of $\tilde{\epsilon}$}
\label{App:eps_tilde}

\begin{table}
	\centering
	\caption{The setting of the stellar evolution code \textsc{mesa} in the calculation of this paper.
	}
	\label{tab:mesa_setting}
	\renewcommand{\arraystretch}{1.4}
	\begin{tabular}{ll} 
		\hline
		Control name & Our setting \\ 
		\hline
		$\mathtt{initial\_mass}$ & $1.4$, $1.5$, $1.6$ or $1.8 \, \mathrm{M}_{\odot}$ \\
		$\mathtt{initial\_Z}$ & $Z=0.02$  \\ 
		$\mathtt{use\_Type2\_opacities}$ & True  \\
		$\mathtt{default\_net\_name}$ & 'basic.net'   \\
		$\mathtt{atm\_T\_tau\_relation}$ & 'Eddington' \\
		$\mathtt{mixing\_length\_alpha}$ & $\alpha_\mathrm{MLT}=1.8$ \\
        $\mathtt{MLT\_option}$ & 'Henyey' \\
		$\mathtt{do\_element\_diffusion}$ & True \\
		$\mathtt{mesh\_delta\_coeff}$ & 0.25 \\
		$\mathtt{max\_years\_for\_timestep}$ & $10^6\,\mathrm{yrs}$ \\
		$\mathtt{diffusion\_rtol\_for\_isolve}$ & $10^{-5}$ \\
		$\mathtt{diffusion\_atol\_for\_isolve}$ & $10^{-6}$ \\
		$\mathtt{use\_Ledoux\_criterion}$ & 
		False or True 
		\\
		$\mathtt{do\_conv\_premix}$ & True \\
		\hline
	\end{tabular}
\end{table}

\begin{figure}
	\includegraphics[width=0.9\columnwidth]{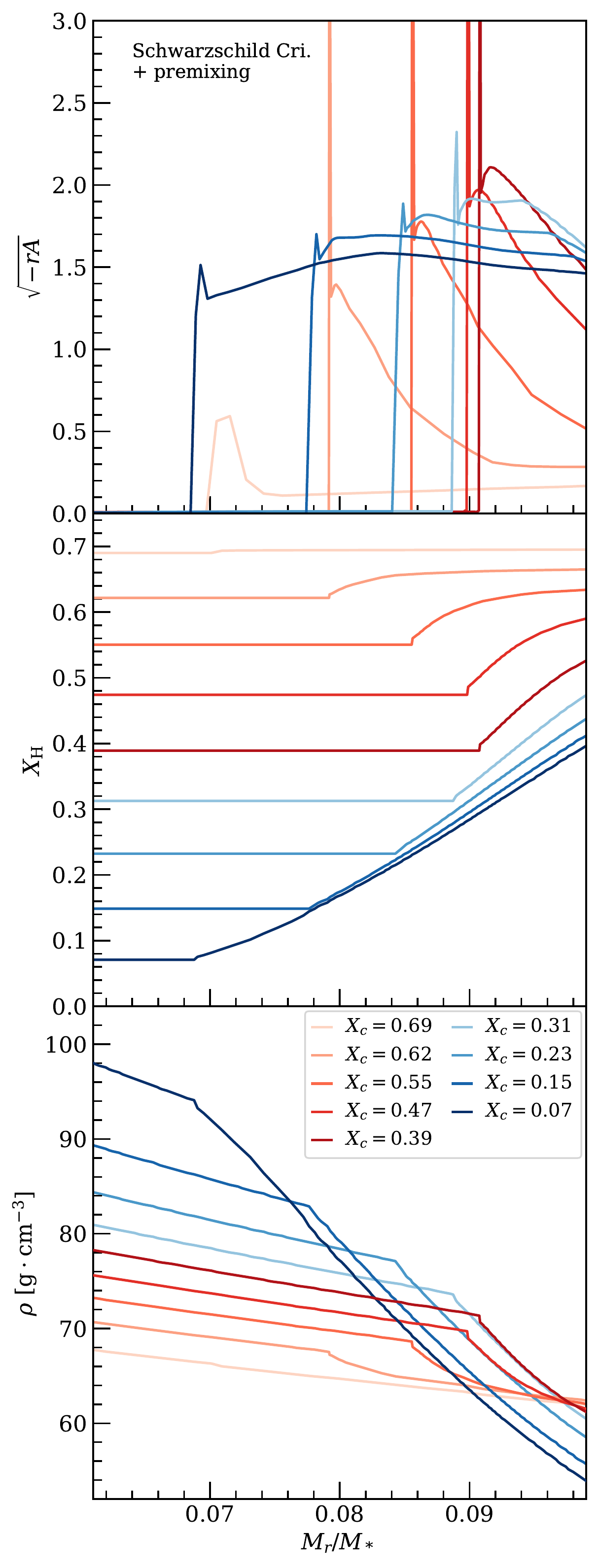}
    \caption{
    Profiles of the normalised Brunt--V\"ais\"al\"a frequency $\sqrt{-rA} \, (\equiv \sqrt{N^2 r/g})$ (upper panel), the hydrogen mass fraction $X_\mathrm{H}$ (middle panel) and the density $\rho$ (lower panel). These are computed by \textsc{mesa}
    with the mass of $1.5\,\textrm{M}_{\odot}$ for 
    various values of the central hydrogen mass fraction, $X_c$,
    which is equal to $0.70$ at the zero-age main-sequence (ZAMS) stage. 
    The abscissa means the concentric mass normalised by the stellar mass $M_*$. The lines with reddish colours (for $X_c \ge 0.40$) and bluish colours (for $X_c \le 0.31$) correspond to the phases of core growth and shrinkage, respectively.}
    \label{fig:N2profile}
\end{figure}

\begin{figure}
	\includegraphics[width=\columnwidth]{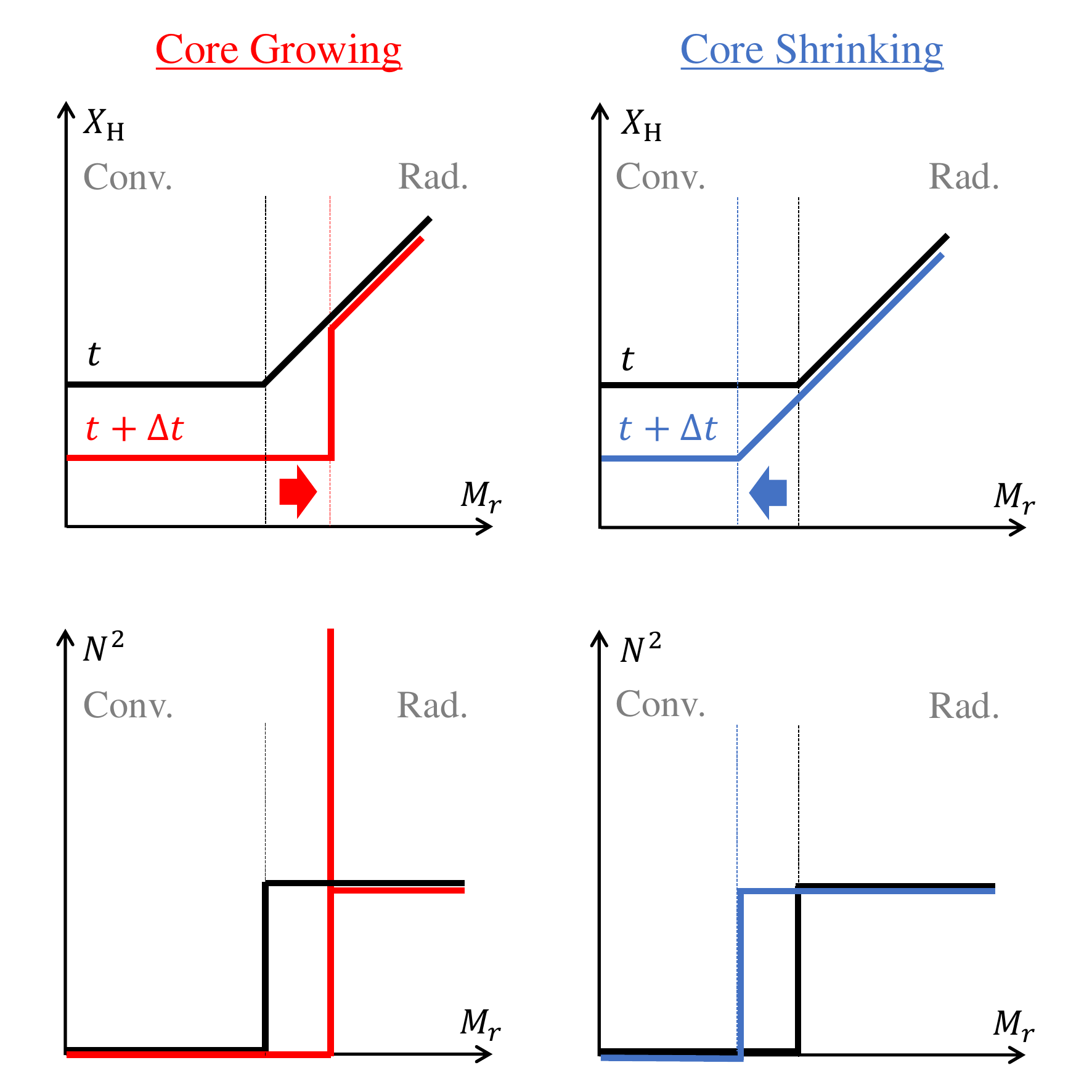}
    \caption{
    Schematic pictures of the evolution of $X_\mathrm{H}$ and $N^2$ profiles during the phases of
    core growing (left panels) 
    and shrinking (right panels). 
    The abscissa represents the concentric mass, while the ordinate means $X_\mathrm{H}$ and $N^2$ in the upper and lower panels, respectively. The vertical dashed lines are the boundary between the
    convective and radiative regions.
    }
    \label{fig:Schematic_XHevo}
\end{figure}

We construct evolutionary models of main-sequence stars with the solar composition by the \textsc{mesa} evolution code (r12778). Our settings of major parameters
are listed in Table \ref{tab:mesa_setting}. As for the convective mixing, we try the following two cases:
(1) $\mathtt{use\_Ledoux\_criterion}$ and $\mathtt{do\_conv\_premix}$ are set to false and true, respectively \citep[cf.][]{paxton2019modules};
(2) the both parameters are set to true.
We refer to the first and second cases as the Schwarzschild case and the Ledoux case, respectively.
Note that, in the both cases, the neutrality condition of the Schwarzschild criterion for convection is approximately satisfied in the radiative layers just outside the convective core because of the convective premixing. The other parameters are set following the test-suite \texttt{1.5M\_with\_diffusion}, which includes element diffusion from gravitational settling and chemical diffusion.

In Fig.~\ref{fig:N2profile}, we plot 
for the Schwarzschild case the profiles of (normalised) Brunt-V\"ais\"al\"a frequency 
$\sqrt{N^2 r/g}$,
hydrogen mass fraction $X_\mathrm{H}$ and density $\rho$ for different values of hydrogen mass fraction at the centre $X_c$.
Since $N^2$ increases very steeply from essentially zero
to a large positive value at the inner edge of the
radiative envelope, we can easily identify
the evolutionary change in
the mass of the convective core 
($M_{\mathrm{core}}$)
from these profiles.
The non-monotonic change of $M_{\mathrm{core}}$ 
for this mass range
is discussed in detail in e.g.\ Section 3.2.3 of \cite{aerts2010asteroseismology}.
The initial increase for $X_c \gtrsim 0.40$
is associated with the decreasing contribution
of the pp chain to the core nuclear burning
compared to the CNO cycle. On the other hand, the 
decrease of $M_{\mathrm{core}}$ in the later phase 
(for $X_c \lesssim 0.40$) is due to the gradual depletion of hydrogen, which brings about the opacity decrease, in the convective core.

Fig.~\ref{fig:N2profile} shows that 
$X_\mathrm{H}$ and $\rho$ are 
essentially
discontinuous
at the boundary between the convective core
and the radiative envelope
when the core grows, while they 
are continuous when the core shrinks.
This difference can be understood
schematically in Fig.~\ref{fig:Schematic_XHevo}.
As shown in 
the top left panel,
$X_{\mathrm{H}}$ decreases in the core because of the nuclear burning near the centre
and is homogenised to the average value due to the convective mixing.
Because
$X_{\mathrm{H}}$ hardly changes outside the core,
it becomes discontinuous at the core boundary.
When the core shrinks
as in the top right panel,
$X_{\mathrm{H}}$ at the outer edge of the convective core
at one time
becomes that at the inner edge of the radiative envelope
at the next time step.
Thus,
$X_{\mathrm{H}}$ near the inner edge of the radiative envelope
essentially records the core values in the past.
It is therefore continuous.
The corresponding profiles of $N^2$ are shown
in the bottom panels.
The most conspicuous difference between the two panels
is that
a sharp spike appears at the boundary
only when the core grows
(bottom left panel)
because $N^2$ depends on the 
spacial
derivative of $X_{\mathrm{H}}$ in the radial direction.

In any case, the 
scale height of the structure
at the boundary
is found to be so short 
compared to the wavelength of oscillations
in the radiative side of the boundary
that the derivative of
$\rho$ can be regarded as
discontinuous at the boundary
(cf. Appendix \ref{App:C}).
We therefore apply the results
of the discontinuous case
in Section \ref{subsec:three_five}
to compute $\tilde{\epsilon}$,
which is defined by
equation
(\ref{tildeepsilon}).
%%%

\section{dependence 
on mesh numbers }
\label{App:C}

\subsection{Comparison between
the wavelength and the scale height}
In Section \ref{sec:three},
we developed two different types of analysis,
that for continuous density profiles 
and that for discontinuous ones.
In order to find which type should be adopted,
we compare the wavelength of oscillations
with the scale height of the structure.

\begin{figure}
	\includegraphics[width=0.9\columnwidth]{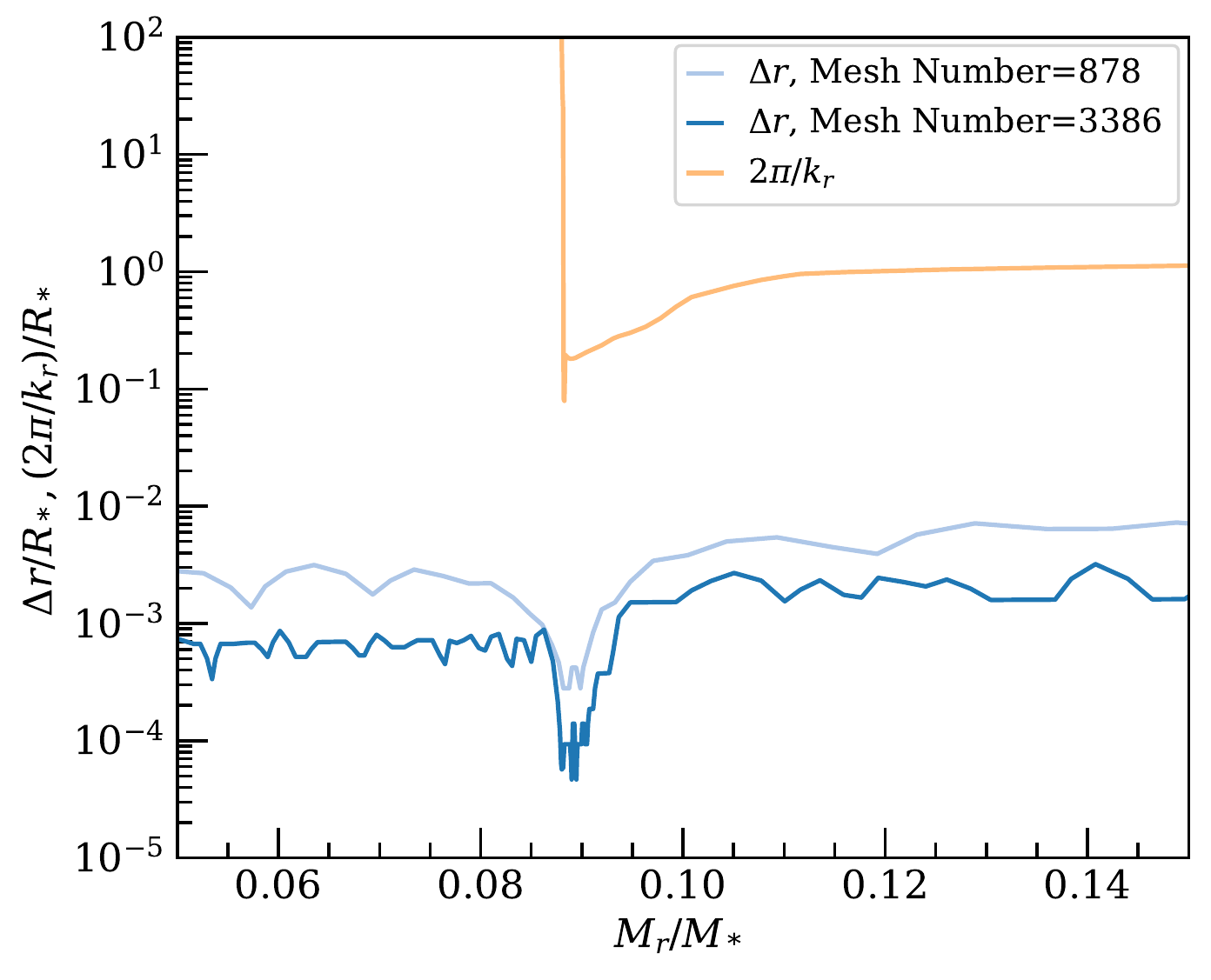}
    \caption{
    Comparison between the mesh size $\Delta r$ (blue lines)
    and the wavelength $2\pi / k_r$
    (orange line) computed by \textsc{mesa} with the mass of $1.5\,\textrm{M}_{\odot}$ as functions of the fractional mass ($M_r/M_*$). The vertical axis is normalised by the stellar radius.
    The mesh size is plotted
    for the two models with different
    total mesh numbers of 869 (deep blue line)
    and 3375 (light blue line).
    Both models have $X_c = 0.51$,
    which corresponds to the age of 
    $0.80 \, \mathrm{Gyr}$.
    The wavelength is computed
    for the finer-mesh model
    with the spin parameter of $s=10$
    and the rotation period of $0.455 \, \mathrm{d}$.
    }
    \label{fig:dr_kr_reci}
\end{figure}

Since the Brunt--V\"ais\"al\"a frequency, $N$,
is the most important for the analysis,
we pay attention to $N$, which changes from
$0$ in the outermost mesh point 
at radius $r = r_c$
in the convective
core
to a large value, $N_1$, in the innermost mesh point at $r = r_r$
in the radiative envelope.
We may therefore estimate the scale height
as
\begin{equation}
N_1 \times \frac{r_r - r_c}{N_1 - 0}
= \Delta r
,
\end{equation}
where $\Delta r = r_r - r_c$ is the radius 
difference between the two mesh points.
We thus adopt the mesh size $\Delta r$
as the scale height of the structure
in the radiative layer
just outside the convective core.
Fig.~\ref{fig:dr_kr_reci}
shows the comparison between the mesh size and the wavelength for two typical models with
different total mesh numbers of 869 and 3375.
The model with the coarse mesh is obtained with
the default settings of \textsc{mesa},
while the fine-mesh model is constructed
with the settings described in Appendix \ref{App:eps_tilde}.

From this figure, we confirm
that the wavelength is much longer in the radiative envelope
by about two orders of magnitude or more
irrespectively of the total mesh number.
We should therefore
adopt the case of discontinuous profiles at the core boundary.
In addition, we find that
the gravito-inertial waves
are well resolved in the radiative envelope
even in the coarse-mesh model.

\subsection{%
Right-hand side limit
of the Brunt--V\"ais\"al\"a frequency
}

In order to calculate $\tilde{\epsilon}$ from equation (\ref{tildeepsilon}), we need to calculate $N_0$, which is the right-hand limit of Brunt--V\"ais\"al\"a frequency 
at the convective-core boundary (see equation \ref{Nb+}). For simplicity, when a sharp spike appears at the boundary in the $N$ profile (cf. the bottom left panel of Fig.~\ref{fig:Schematic_XHevo}), we choose 
the value at the next mesh point 
on the right-hand side of 
the spike as the right-hand limit. 
We identify the spike
if $N^2$ has
a local maximum 
that
is twice or more larger than the values
at the next mesh points on
both sides. On the other hand, when no spike is located (cf. the bottom 
right panel of Fig.~\ref{fig:Schematic_XHevo}), we 
adopt the local maximum of $N$
around the boundary
as the right-hand side limit.

Note that the intermediate convection zones 
sometimes arise in the layers of 
chemical composition gradients
when the core grows in the \textsc{mesa} calculation particularly
if the Schwarzschild criterion 
for convection is adopted. 
In that case, we  regard the outer boundary
of the outermost intermediate
convection zone as the core boundary
in the above method.

\begin{figure}
	\includegraphics[width=0.9\columnwidth]{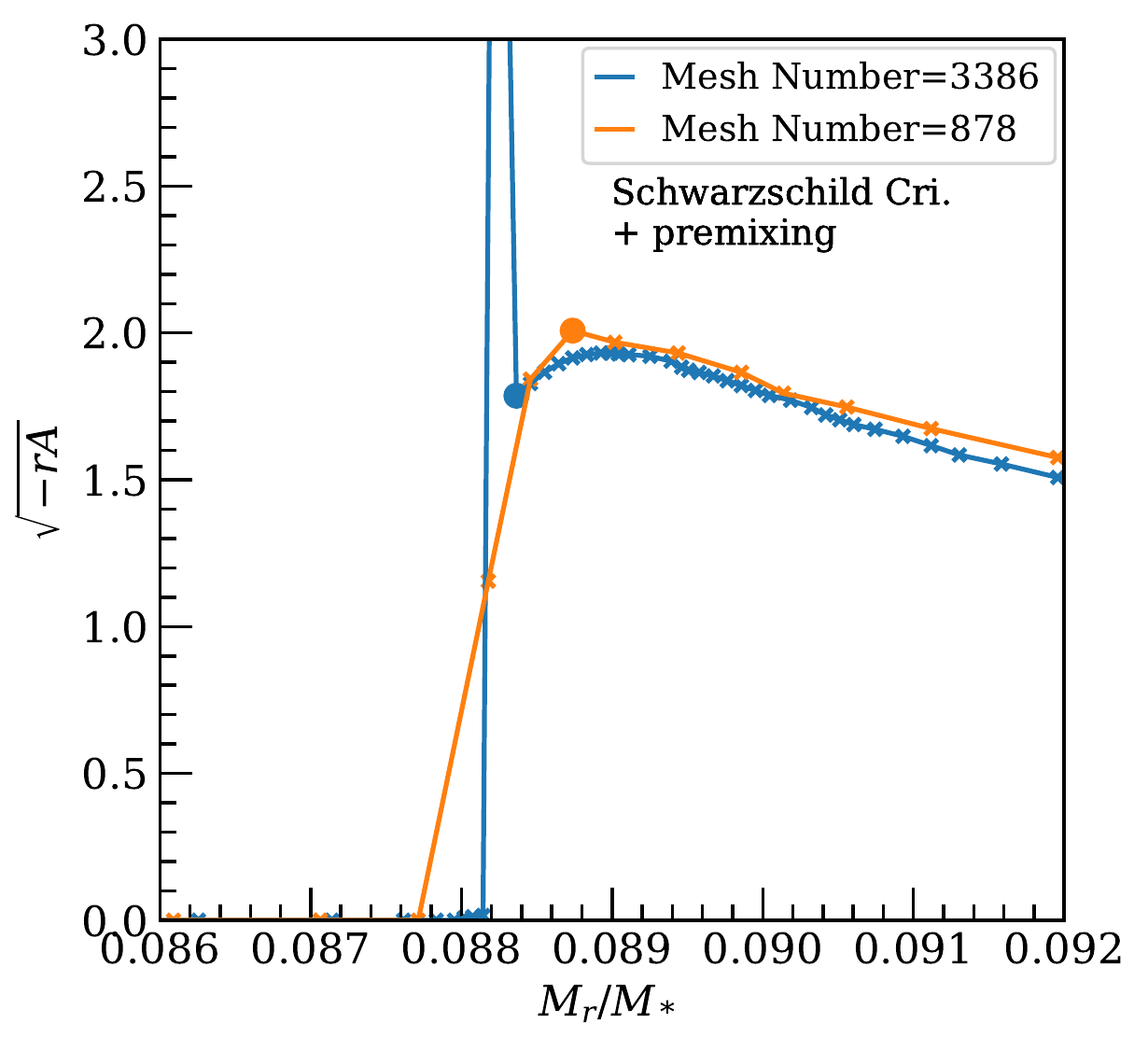}
    \caption{
    Profiles of the normalised Brunt--V\"ais\"al\"a frequency 
    $\sqrt{-rA} = \sqrt{r/g} N$
    as functions of the fractional mass ($M_r/M_*$) computed by \textsc{mesa} with the mass of $1.5\,\textrm{M}_{\odot}$ 
    and $X_c = 0.51$ (which corresponds to the age of $0.80 \, \mathrm{Gyr}$ ).
    The blue and orange lines correspond to the cases of the total mesh number of 
    3388 and 878, respectively. The small 
    crosses represent the positions of mesh points,
    while the large filled circles indicate the mesh points
    at which the 
    right-hand limits of $N$ at the core boundary are estimated.
    }
    \label{fig:N2profile_compare_mesh}
\end{figure}

Fig.~\ref{fig:N2profile_compare_mesh} shows the profiles of normalised Brunt-Väisälä frequency ($\sqrt{r/g} N$) for two
different spatial mesh numbers 
with
the estimated right-hand side limits
at the core boundary. 
We regard that the model with
the mesh number of 3388 has a spike,
while that with 878 does not.
The reason for no spike
in the latter model is that
the spatial resolution
is too low to resolve 
the rapid change in the gradient
of chemical compositions 
around the core boundary.
In spite of this difference,
there is
little influence
on the estimate of $N_0$
in these particular examples.

\begin{figure}
	\includegraphics[width=0.9\columnwidth]{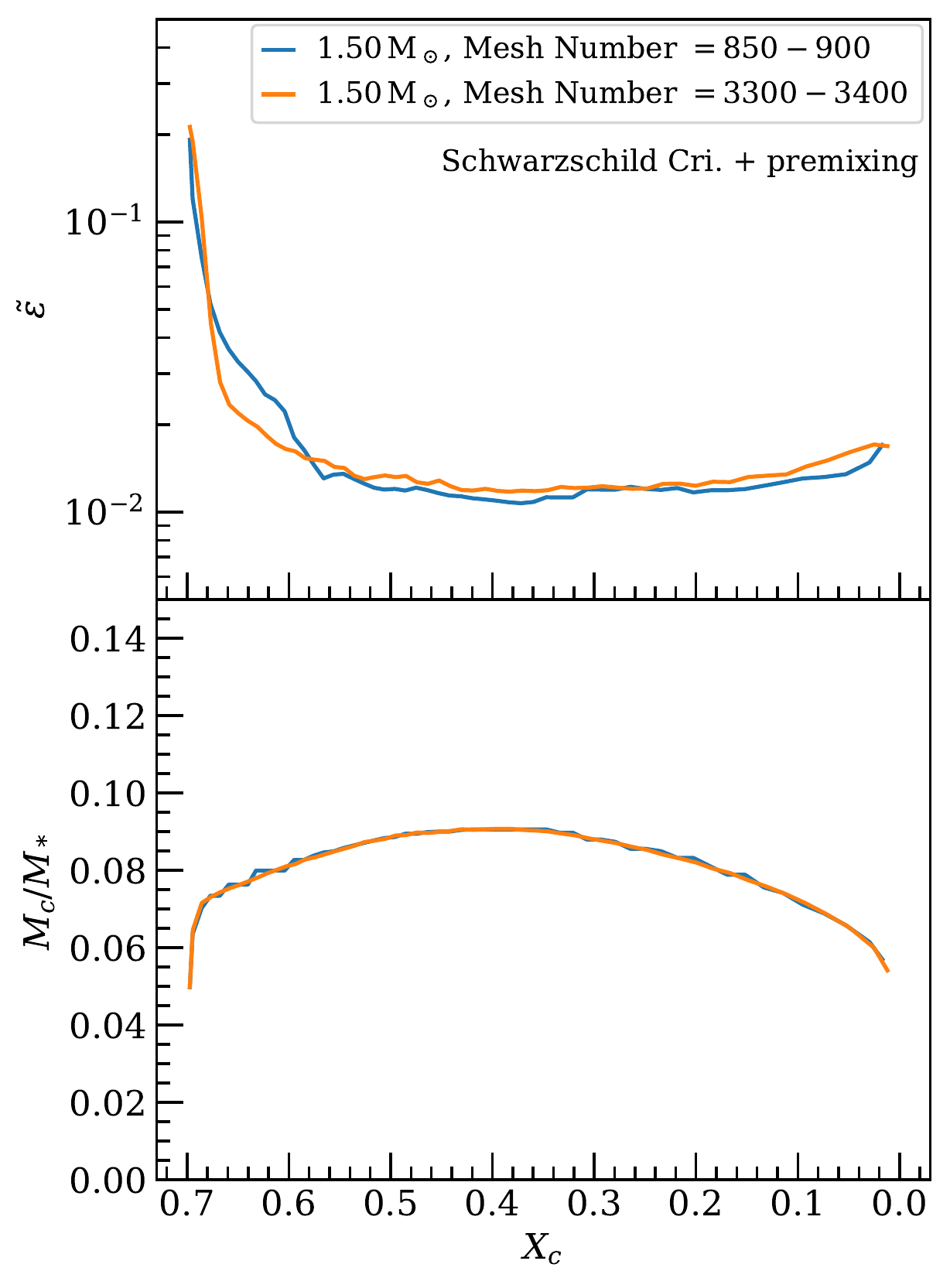}
    \caption{
    Evolution of $\tilde{\epsilon}$ and $M_\mathrm{core}$ as function of
    the central hydrogen mass fraction $X_c$ for the models with $1.5 \, \mathrm{M_\odot}$. We set the rotation period to $0.455 \, \mathrm{d}$. The ordinates represent $\tilde{\epsilon}$ (upper panels) and $M_\mathrm{core}$ normalised by the total stellar mass $M_*$
    (lower panels). The blue and orange lines correspond to the models with
    coarse and fine mesh points, respectively (refer to the main text for the details).
    In the both cases,
    the Schwarzschild criterion for convection is adopted,
    and the convective premixing is
    taken into account.
    Note that $X_c$ is equal to $0.70$
    at the zero-age main-sequence stage and
    decreases during the evolution afterwards. 
    }
    \label{fig:eps_Mcore_evo_compare_mesh}
\end{figure}

On the other hand, Fig.~\ref{fig:eps_Mcore_evo_compare_mesh}
shows the evolution of $\tilde{\epsilon}$ and $M_\mathrm{core}$ for two different settings
about the mesh numbers.
The blue lines are obtained from
the coarse-mesh models
with the default settings of 
\textsc{mesa},
whereas
the orange lines correspond
to the fine-mesh models
constructed
with the settings described in
Appendix \ref{App:eps_tilde}.
There is little difference in
$M_{\mathrm{core}}$ between the
two cases,
whereas a distinct discrepancy 
in $\tilde{\epsilon}$ is found 
for $0.58 \lesssim X_c \lesssim 0.68$.
The discrepancy is also because of the
too poor spatial resolution of the
coarse-mesh models
to resolve the variation
in chemical compositions.
These examples demonstrate
the importance
of having
sufficient mesh points in models
to compute $\tilde{\epsilon}$ 
precisely.

%%%%%%%%%%%%%%%%%%%%%%%%%%%%%%%%%%%%%%%%%%%%%%%%%%

% Don't change these lines
\bsp	% typesetting comment
\label{lastpage}
\end{document}